\documentclass[%
 reprint,
 amsmath,amssymb,
 aps,
]{revtex4-1}
\usepackage{amsmath,amssymb}
\usepackage{amsfonts}
\usepackage[letterpaper, portrait, margin=3cm]{geometry}
\usepackage{graphicx}
\usepackage{dcolumn}
\usepackage{bm}

\usepackage[labelfont={sf,bf}, textfont=sf,hangindent=18pt,justification=raggedright]{caption,subfig}
\usepackage{physics}
\usepackage{multirow}
\newcommand{\dfourx}{\mathrm{d}^4 x}

\newcommand{\dt}{\mathrm{d}t}
\renewcommand{\Im}{\mathrm{Im}}
\newcommand{\as}{\alpha_s}

\newcommand{\quarkthreed}{\langle\overline{q}q\rangle}
\newcommand{\mquarkthreed}{\langle m\overline{q}q\rangle}
\newcommand{\gluonfourd}{\langle\alpha G^2\rangle}
\newcommand{\mixed}{\langle g\overline{q} \sigma G q\rangle}
\newcommand{\gluonsixd}{\langle g^3 G^3\rangle}
\newcommand{\quarksixd}{\langle\overline{q}q\overline{q}q\rangle}
\newcommand{\vevomega}[1]{\langle\Omega|#1|\Omega\rangle}
\newcommand{\angled}[1]{\langle #1\rangle}

\newcommand{\lsr}{\mathcal{R}}
\newcommand{\gev}{\ensuremath{\text{GeV}}}
\newcommand{\mev}{\ensuremath{\text{MeV}}}

\begin{document}

\preprint{1806.02465}

\title{Mass Calculations of Light Quarkonium, Exotic $J^{PC}=0^{+-}$ Hybrid Mesons 
from Gaussian Sum-Rules}
\author{J.~Ho}
\email{j.ho@usask.ca}
\affiliation{Department of Physics and Engineering Physics,\\ University of Saskatchewan,\\           Saskatoon, SK, S7N 5E2, Canada}
\author{R.~Berg}
\email{rtb867@mail.usask.ca}
\affiliation{Department of Physics and Engineering Physics,\\ University of Saskatchewan,\\           Saskatoon, SK, S7N 5E2, Canada}
\author{Wei Chen}
\email{chenwei29@mail.sysu.edu.cn}
\affiliation{School of Physics,\\ Sun Yat-Sen University,\\ Guangzhou 510275, China}
\author{D.~Harnett}
\email{derek.harnett@ufv.ca}
\affiliation{Department of Physics,\\ University of the Fraser Valley,\\ Abbotsford, BC, V2S 7M8, Canada}
\author{T.G.~Steele}
\email{tom.steele@usask.ca}
\affiliation{Department of Physics and Engineering Physics,\\ University of Saskatchewan,\\           Saskatoon, SK, S7N 5E2, Canada}

\date{\today}

\begin{abstract}
We extend previous calculations of leading-order correlation functions of spin-0 and 
spin-1 light quarkonium hybrids to include QCD condensates of dimensions five and six, 
with a view  to improving the stability of QCD sum-rules analyses in 
previously unstable channels.  
Based on these calculations, prior analyses in the literature, 
and its  experimental importance,
we identify the exotic $J^{PC}=0^{+-}$ channel as the most promising for detailed study.
Using Gaussian sum-rules constrained by the H\"older inequality, we calculate masses 
of light (nonstrange and strange) quarkonium hybrid mesons with $J^{PC}=0^{+-}$.
A model-independent analysis of the hadronic spectral function
indicates that there is distributed resonance strength in this
channel. 
Hence, we study two hadronic models with distributed resonance strength:
a single wide resonance model and a double narrow resonance model.
The single wide resonance model is disfavoured as it leads to an anomalously large resonance width (greater than 1~GeV).
The double narrow resonance model yields
excellent agreement between QCD and phenomenology:
in both nonstrange and strange cases, we find hybrid masses of $2.60$~GeV and $3.57$~GeV.

\end{abstract}

\maketitle


\section{Introduction}\label{I}

\noindent It has long been conjectured that hadrons could exist beyond the conventional quark model of quark-antiquark ($q\bar q$) mesons and three-quark ($qqq$, $\bar q\bar q\bar q$) baryons. 
In particular, colour-singlet hybrid mesons consisting of a quark, antiquark, and explicit gluonic degree of freedom have a long history~\cite{Jaffe1976}.
While evidence of hadronic structures outside of the conventional model has been accumulating with experimental observations and confirmations of tetraquarks~\cite{Aaij2014,Liu2013,Ablikim2013} and pentaquarks~\cite{Aaij2015}, an experimental confirmation of hybrid mesons has eluded observation. 
Designed to search for light hybrid mesons
(particularly those with exotic $J^{PC}$ that do not exist in the conventional quark model), 
the GlueX experiment at Jefferson Lab~\cite{GlueX2017} is currently underway, and is anticipated to  give crucial insight into the existence and structure of light hybrids.

The characterization of light hybrid states within the framework of QCD is important.
Identifying the spectrum of the lightest hybrid supermultiplet ($J^{PC} \in \left\{ 1^{--},\,(0,\,1,\,2)^{-+}\right\}$, where the $q\bar{q}$ are in an $S$-wave configuration) and the neighbouring larger supermultiplet ($J^{PC} \in \left\{ 0^{+-},\,1^{+-},\,2^{+-},\,3^{+-},\,(0,1,2)^{++}\right\}$, where the $q\bar{q}$ are in a $P$-wave configuration) 
 is 
of particular  interest from an experimental perspective, and is aligned with the mandate of the GlueX experiment~\cite{GlueX2017}.
There have been numerous studies done on light quark hybrids covering a range of quantum numbers using 
QCD Laplace sum-rules
(LSRs)~\cite{Balitsky:1982ps,Govaerts:1983ka,Govaerts:1984bk,Latorre:1984kc,Govaerts1985,Balitsky:1986hf,Braun1986,Govaerts:1986pp,Latorre1987,Huang1999,Chetyrkin2000,Jin:2000ek,Jin:2002rw,Narison2009,Feng2014,Huang2015,Huang2016}, 
lattice QCD~\cite{Dudek2011,Dudek2013}, 
the Schwinger-Dyson formalism~\cite{Burden1997,Burden2002,Hilger2017},
the flux tube model~\cite{Close:1994hc,Barnes:1995hc},
and the MIT bag model~\cite{Barnes:1982zs,Chanowitz:1982qj}.
In particular, Reference~\cite{Govaerts1985} contains a comprehensive LSRs analysis of light
hybrids for all $J^{PC}$ with $J\in\{0,\,1\}$ that takes into account condensates up to
dimension four (i.e., 4d).
Analyses of the $0^{++}$, $0^{--}$, $1^{++}$, and $1^{--}$ sectors were stable;
analyses of the $0^{+-}$, $0^{-+}$, $1^{+-}$, and $1^{-+}$ sectors were unstable.
Expected to be the lightest hybrid with exotic quantum numbers, the $1^{-+}$ has been the 
subject of much additional study.
Reference~\cite{Latorre1987} contains a (error-free) $1^{-+}$ hybrid correlator that includes 
condensates up to 6d.  By analyzing lower-weight LSRs than 
 those used in~\cite{Govaerts1985},
the authors arrived at a stable mass prediction.
Subsequently, a variety of improvements (e.g., radiative corrections and higher dimension condensates) 
were included in the $1^{-+}$ hybrid correlator, and the LSRs analyses were updated
accordingly~\cite{Chetyrkin2000,Jin:2000ek,Jin:2002rw,Narison2009,Feng2014,Huang2015}.
In the LSRs analysis of~\cite{Huang1999}, a stable mass prediction for the $0^{-+}$ 
was found using a current different from that of~\cite{Govaerts1985}.
The only stable LSRs analysis of the $0^{+-}$ channel~\cite{Braun1986}
used higher-dimension currents and required estimation of the low-energy theorem term from other channels, introducing multiple sources of theoretical uncertainty.  
Thus, further QCD sum-rules studies of the $0^{+-}$ channel are necessary.

In~\cite{Qiao2012,Berg2012,Chen:2013zia,Chen:2013eha}, it was found that
the inclusion of higher-dimension condensates stabilized previously unstable LSRs analyses from~\cite{Govaerts1985} of hybrids containing heavy quarks.  
Therefore, in Section \ref{II} we provide a systematic computation of leading-order (LO) 5d and 6d condensate
contributions for all light quarkonium hybrids of spin-zero and spin-one. 
Unfortunately, as discussed in Section \ref{III}, these higher-dimension condensates do not stabilize the unstable light hybrid LSRs analyses
as they do for heavy hybrids.
However, in~\cite{Govaerts1985}, it was proposed that the instability in the LSRs  
might be resolved by accounting for finite width effects, an issue also raised 
in~\cite{Braun1986}.
As we show in Section~\ref{V}, a model-independent
analysis of the $0^{+-}$ hadronic spectral function
indicates that there is distributed 
(as opposed to concentrated)
resonance strength in this channel.
To explore width effects and the possibility of excited states, 
we depart from previous LSRs methods. 
Gaussian sum-rules (GSRs)~\cite{Bertlmann1985} are sensitive probes of width effects 
and both ground and excited states, 
and have been shown to be a powerful and versatile analysis 
methodology~\cite{Orlandini2001,Harnett2001,Zhang2003,Harnett2011}.
In particular, the QCD sum-rules paradigm of the $\rho$ meson was used to benchmark and validate these GSR methodologies~\cite{Orlandini2001}.
Thus, in this article, we use GSRs to investigate 
the possibility of distributed resonance strength in
the exotic $0^{+-}$ light hybrid channel. 

In Section~\ref{II}, we calculate LO spin-0 and spin-1 
correlation functions of light quarkonium hybrid currents, including condensates up to 6d. 
Section~\ref{III} includes a review of  the GSRs formalism, and a theoretical constraint on the GSRs based on the H\"older inequality is developed in Section~\ref{IV}.  
The GSRs analysis methodology and results for the $0^{+-}$ channel are presented in Section~\ref{V} with concluding remarks in Section~\ref{VI}.

\section{Hybrid Currents and Correlation Functions}\label{II}
To investigate light quarkonium hybrids, we use currents of the form
\begin{equation}
j_{\mu} = g_{s} \bar{q}\Gamma^{\nu} t^{a} \mathcal{G}_{\mu \nu}^{a}q,
\label{hybridcurrent}
\end{equation}
where $q$ is a light (nonstrange or strange) quark field and
$t^a$ are generators of the fundamental representation of SU(3).
Each combination of
$\mathcal{G}_{\mu \nu}^{a} \in \left\{ G^{a}_{\mu\nu},\,\,\tilde{G}^{a}_{\mu\nu} = \frac{1}{2} \epsilon^{\mu \nu \rho \sigma} G^{a}_{\rho \sigma} \right\}$
and Dirac structure $\Gamma^{\nu}$ together corresponds to particular values of
$J^{PC}$~\cite{Govaerts1985,Latorre1987}; 
these combinations are summarized in Table~\ref{JPC_table}.
\begin{center}
\captionof{table}{The $J^{PC}$ combinations probed through different choices of $\Gamma^{\nu}$ and $\mathcal{G}^a_{\mu \nu}$ in~(\ref{hybridcurrent}).}\label{JPC_table}
\begin{table}[h]
\begin{tabular}{c|c||c}
  $\Gamma^{\nu}$ & $\mathcal{G}^{a}_{\mu\nu}$ & $J^{PC}$\\
  \hline
  $\gamma^{\nu}$ & $G^{a}_{\mu\nu}$ & $0^{++},\,1^{-+}$ \\
  $\gamma^{\nu}$ & $\tilde{G}^{a}_{\mu\nu}$ & $0^{-+},\,1^{++}$ \\
  $\gamma^{\nu}\gamma_5$ & $G^{a}_{\mu\nu}$ & $0^{--},\,1^{+-}$ \\
  $\gamma^{\nu}\gamma_5$ & $\tilde{G}^{a}_{\mu\nu}$ & $0^{+-},\,1^{--}$
\end{tabular}
\end{table}
\end{center}

For each current~\eqref{hybridcurrent}, we calculate and decompose
a diagonal correlation function as follows: 
\begin{align}
  \Pi_{\mu\nu}(q) &= i\int\!\dfourx\, e^{i q\cdot x} 
    \vev{\tau j_{\mu}(x)j^{\dag}_{\nu}(0)}\label{correlator}\\
  &= \frac{q_{\mu}q_{\nu}}{q^2}\Pi^{(0)}(q^2) 
   + \bigg(\frac{q_{\mu}q_{\nu}}{q^2}-g_{\mu\nu}\bigg)\Pi^{(1)}(q^2)\label{spin_breakdown}
\end{align}
where $\Pi^{(0)}$ probes spin-0 states and $\Pi^{(1)}$ probes spin-1 states.

\begin{figure*}[htp!]
\subfloat[Diagram I (LO perturbation theory)]{%
    \includegraphics[width=.26\textwidth]{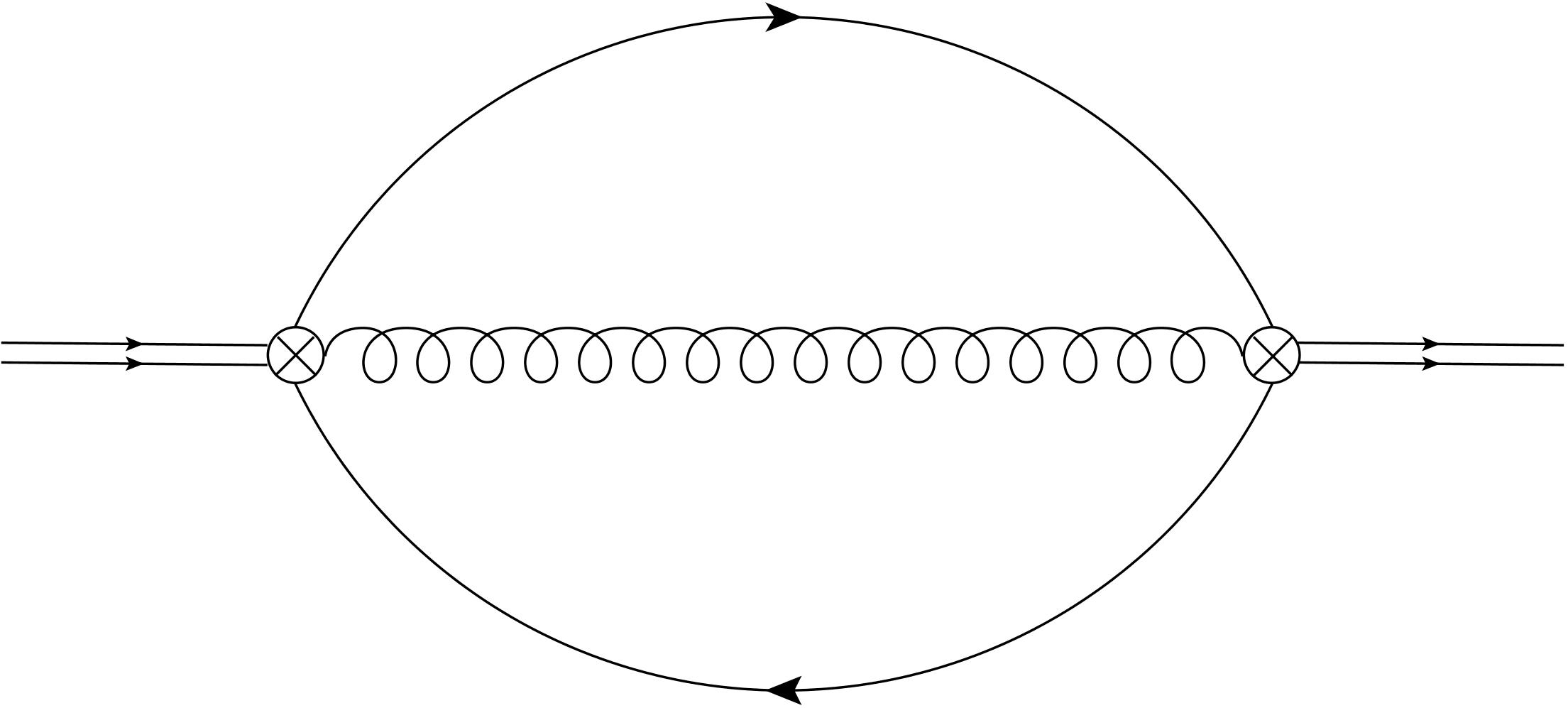}}\hfill
  \subfloat[Diagram II (dimension-three)]{%
    \includegraphics[width=.26\textwidth]{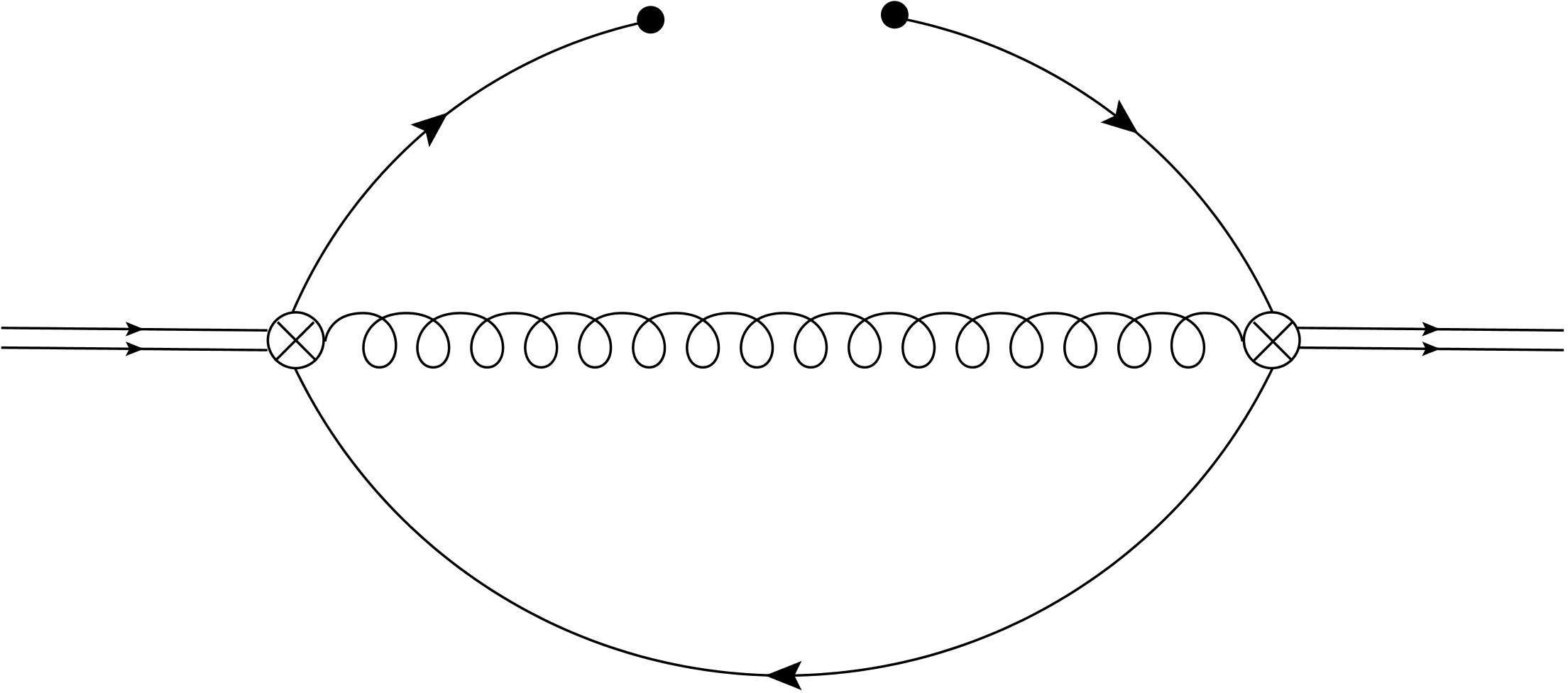}}\hfill
  \subfloat[Diagram III (dimension-four)]{%
    \includegraphics[width=.26\textwidth]{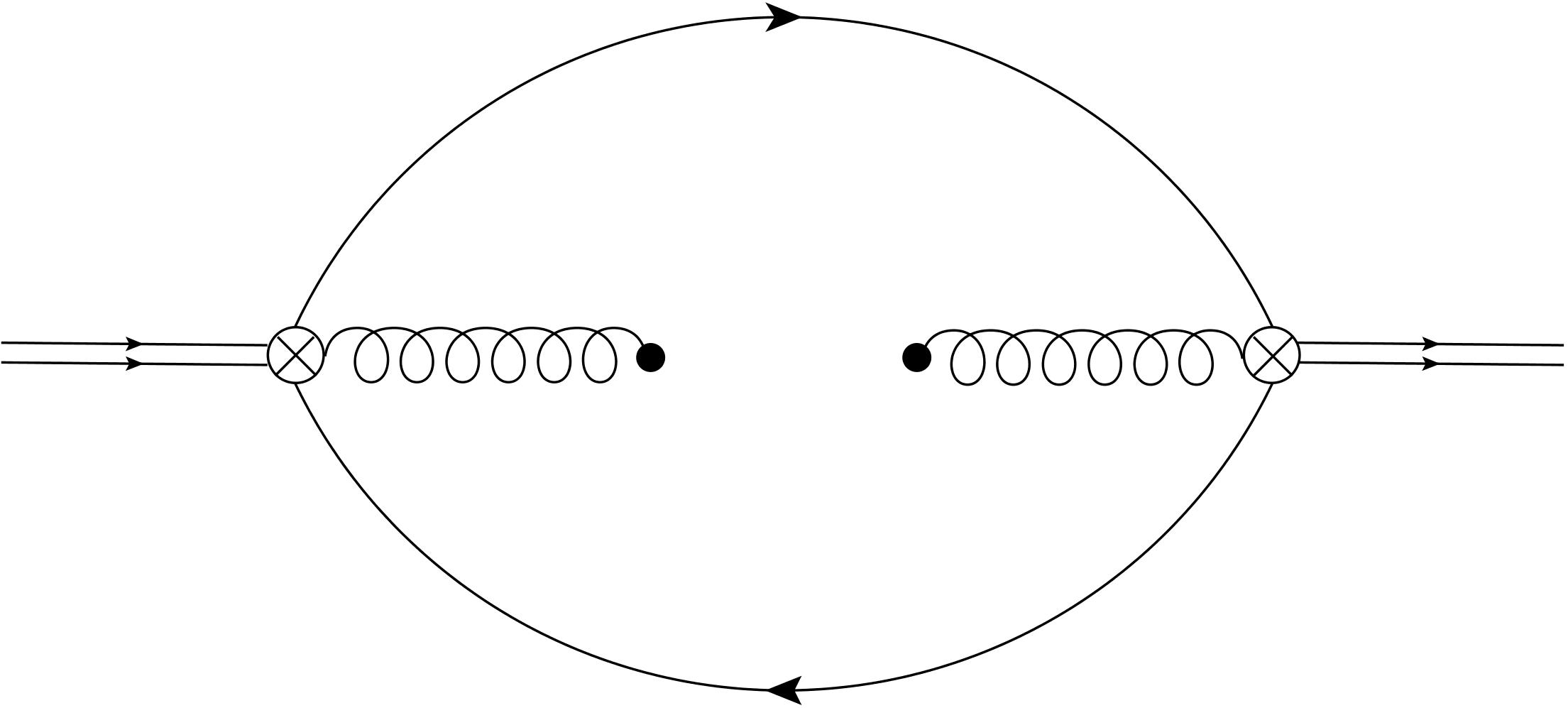}}\\
  \subfloat[Diagram IV (dimension-six)]{%
  	\includegraphics[width=.26\textwidth]{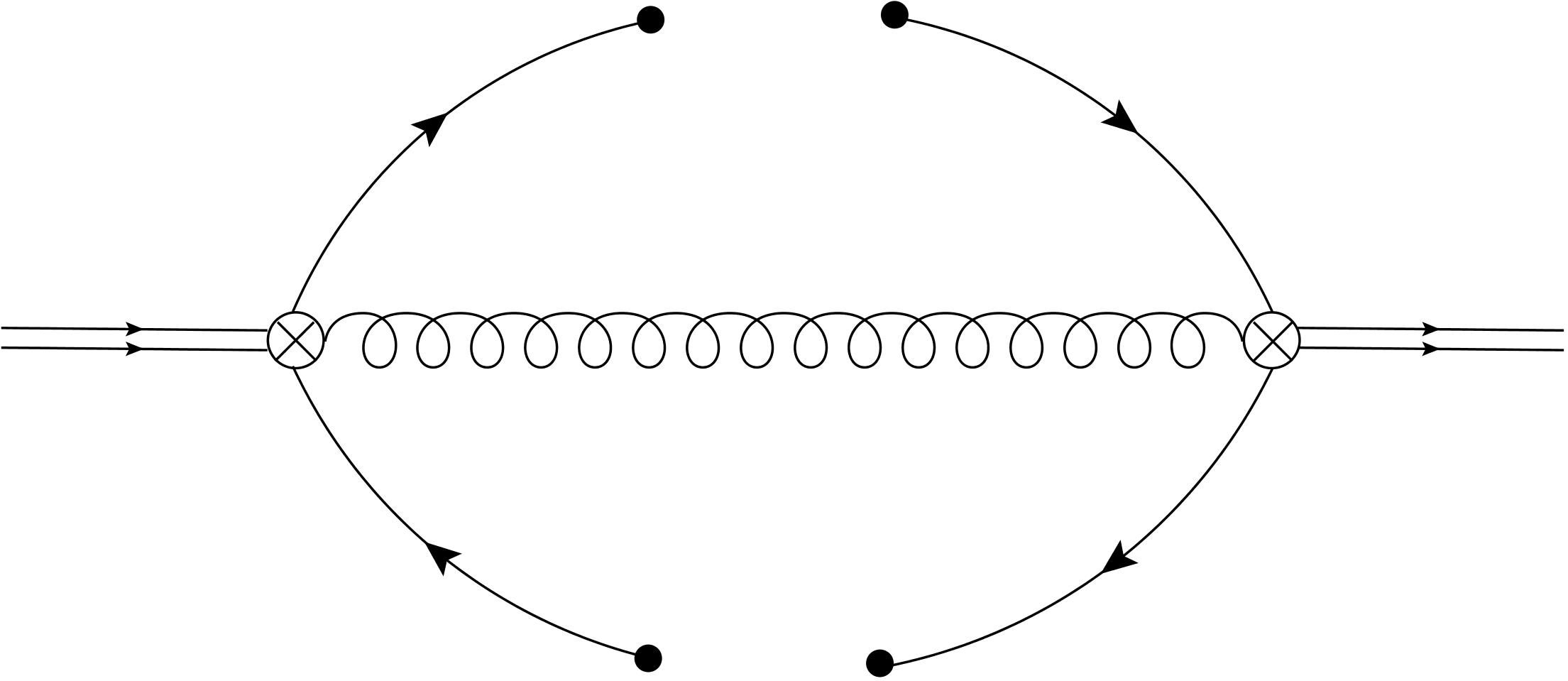}}\hfill
  \subfloat[Diagram V (dimension-six)]{%
    \includegraphics[width=.26\textwidth]{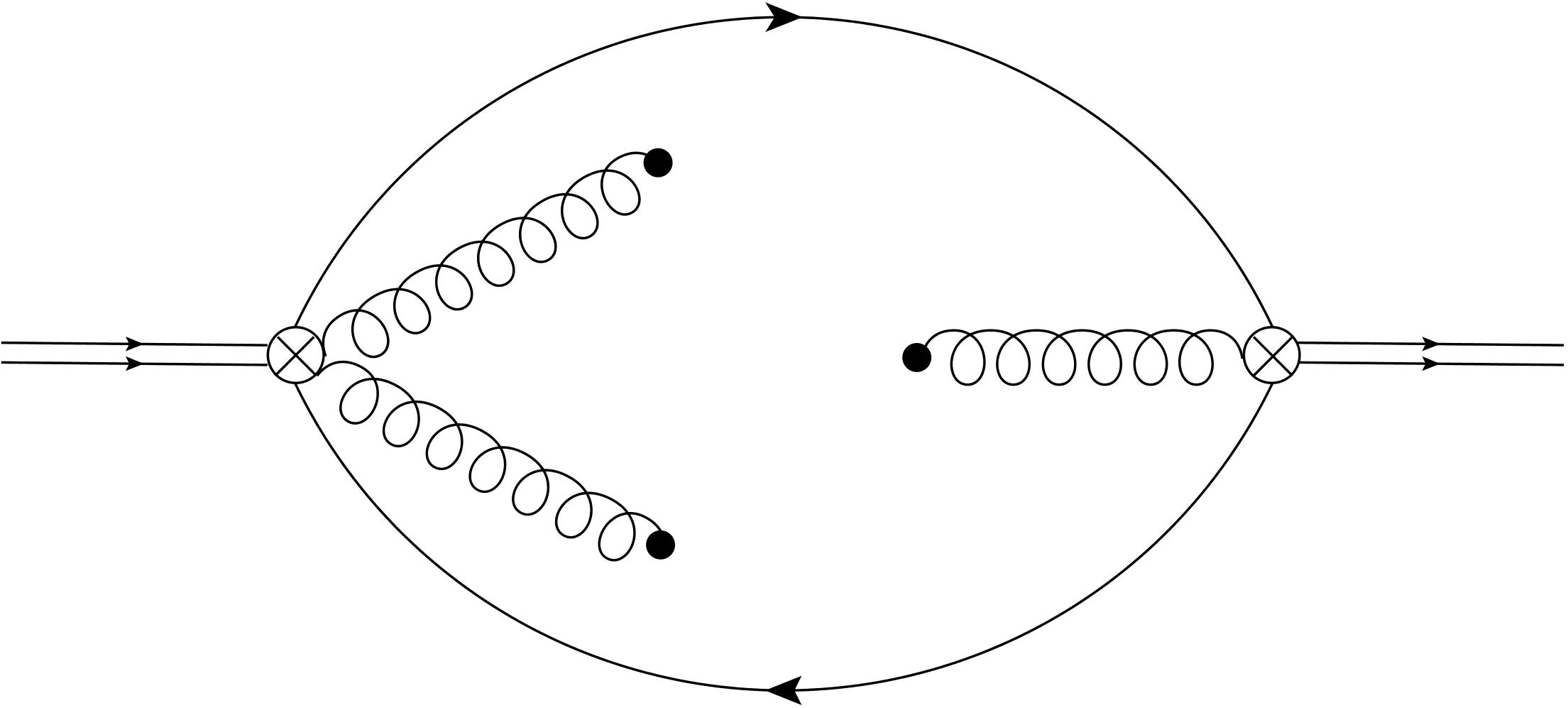}}\hfill
  \subfloat[Diagram VI (dimension-six)]{%
  	\includegraphics[width=.26\textwidth]{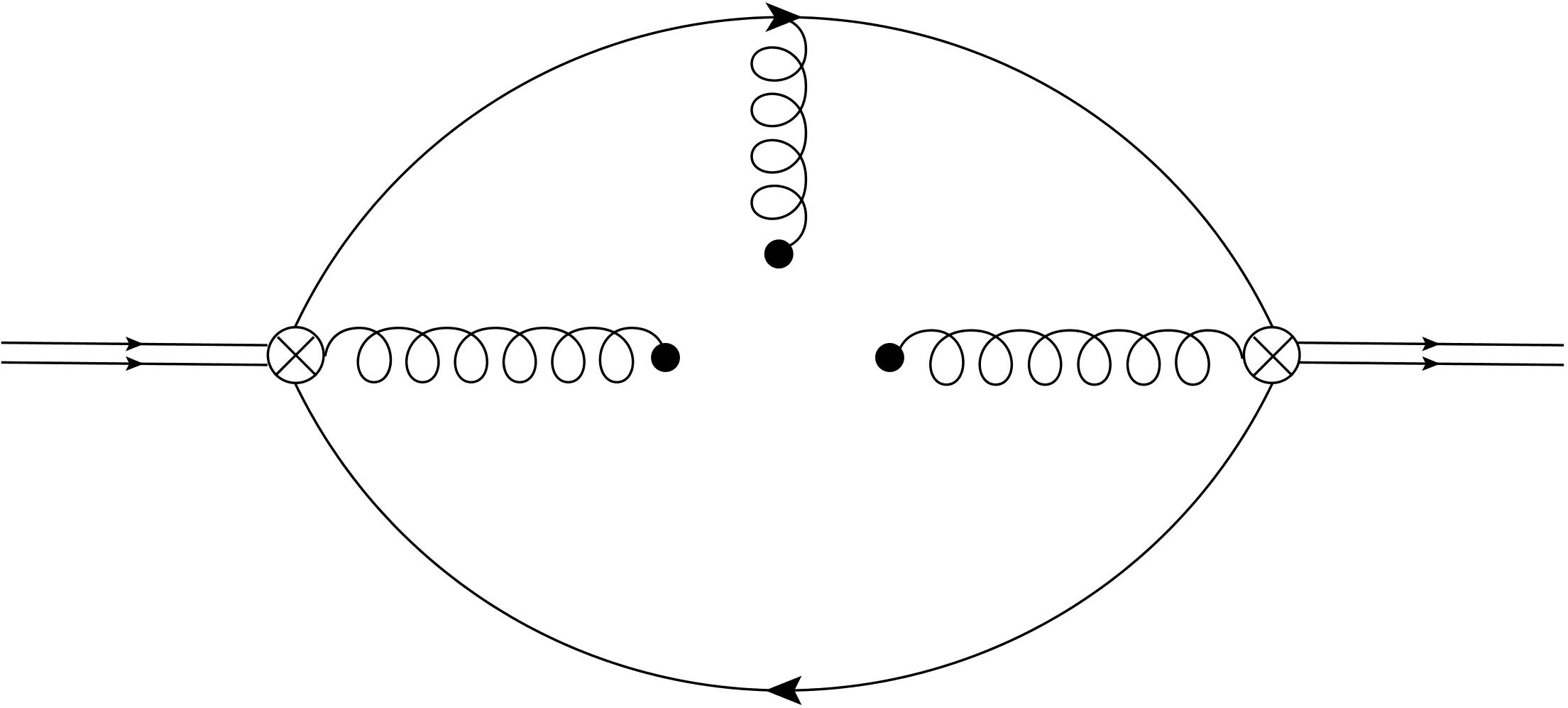}}\\
  \subfloat[Diagram VII (dimension-five)]{%
    \includegraphics[width=.26\textwidth]{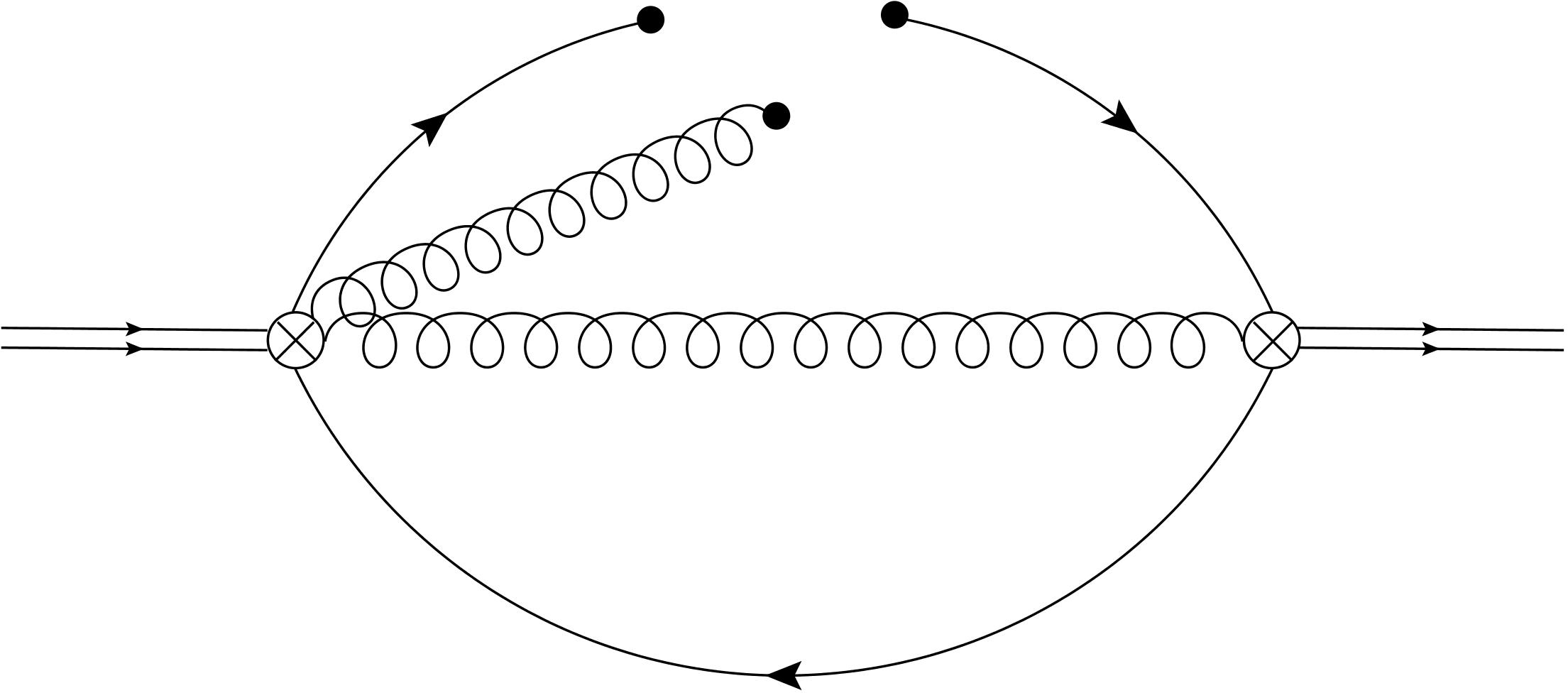}}\hfill
  \subfloat[Diagram VIII (dimension-five)]{%
    \includegraphics[width=.26\textwidth]{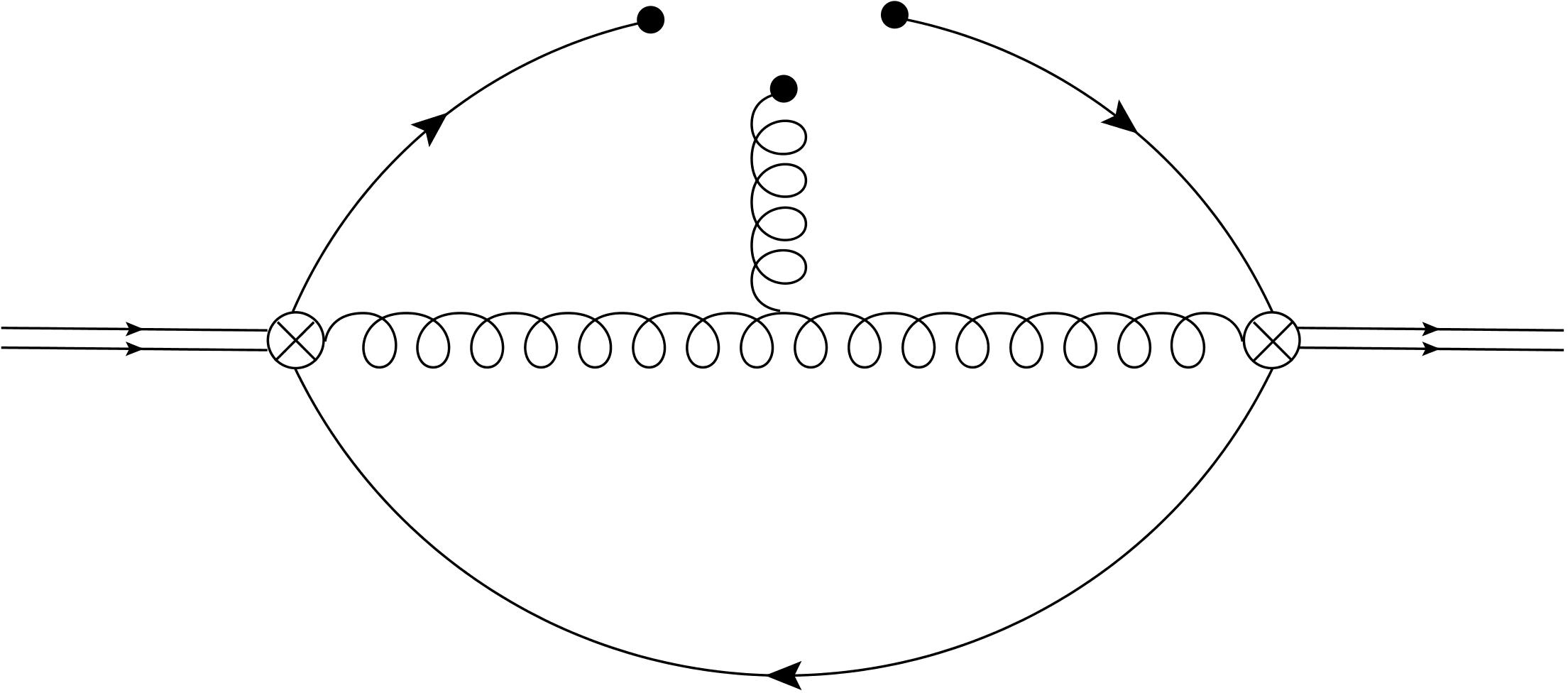}}\hfill
  \subfloat[Diagram IX (dimension-five)]{%
    \includegraphics[width=.26\textwidth]{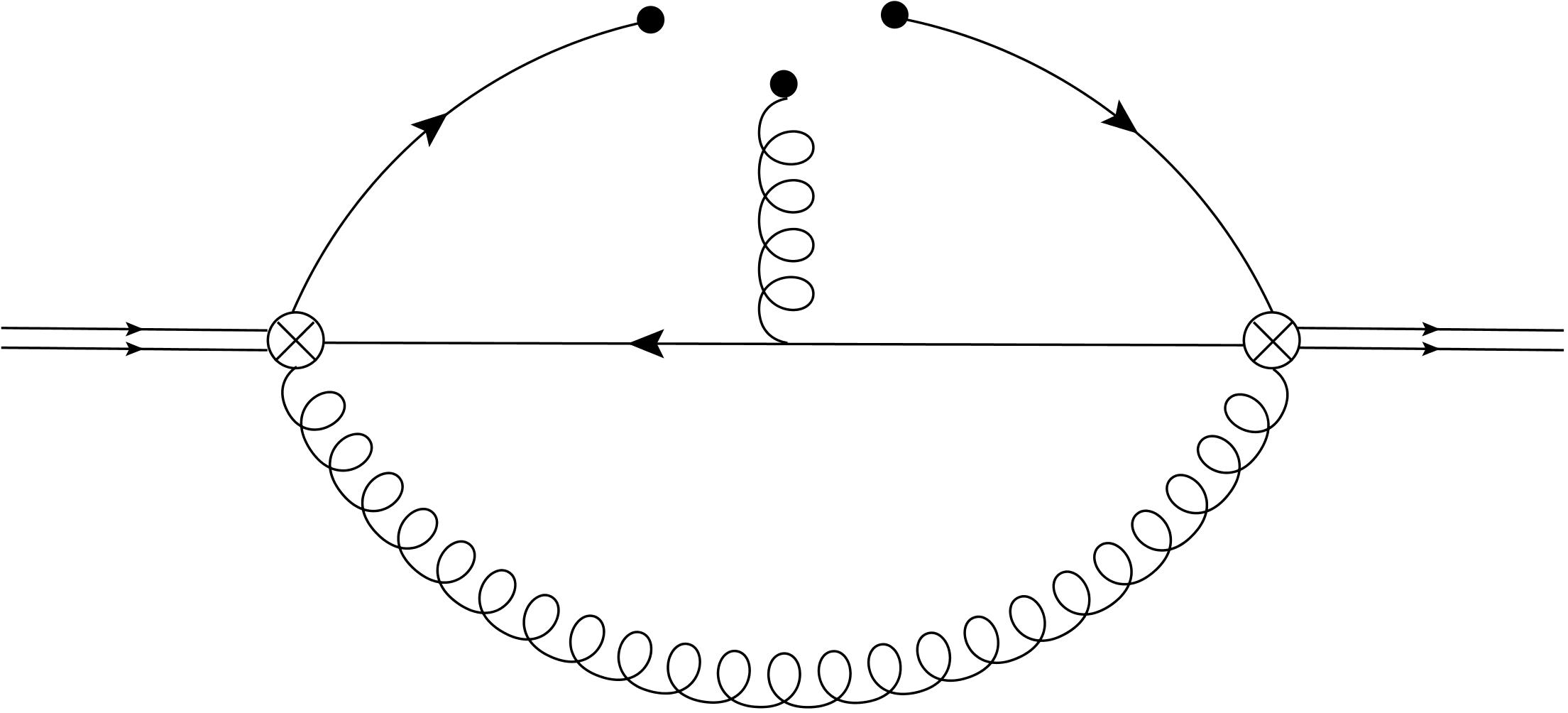}}\\
  \subfloat[Diagram X (dimension-five)]{%
    \includegraphics[width=.26\textwidth]{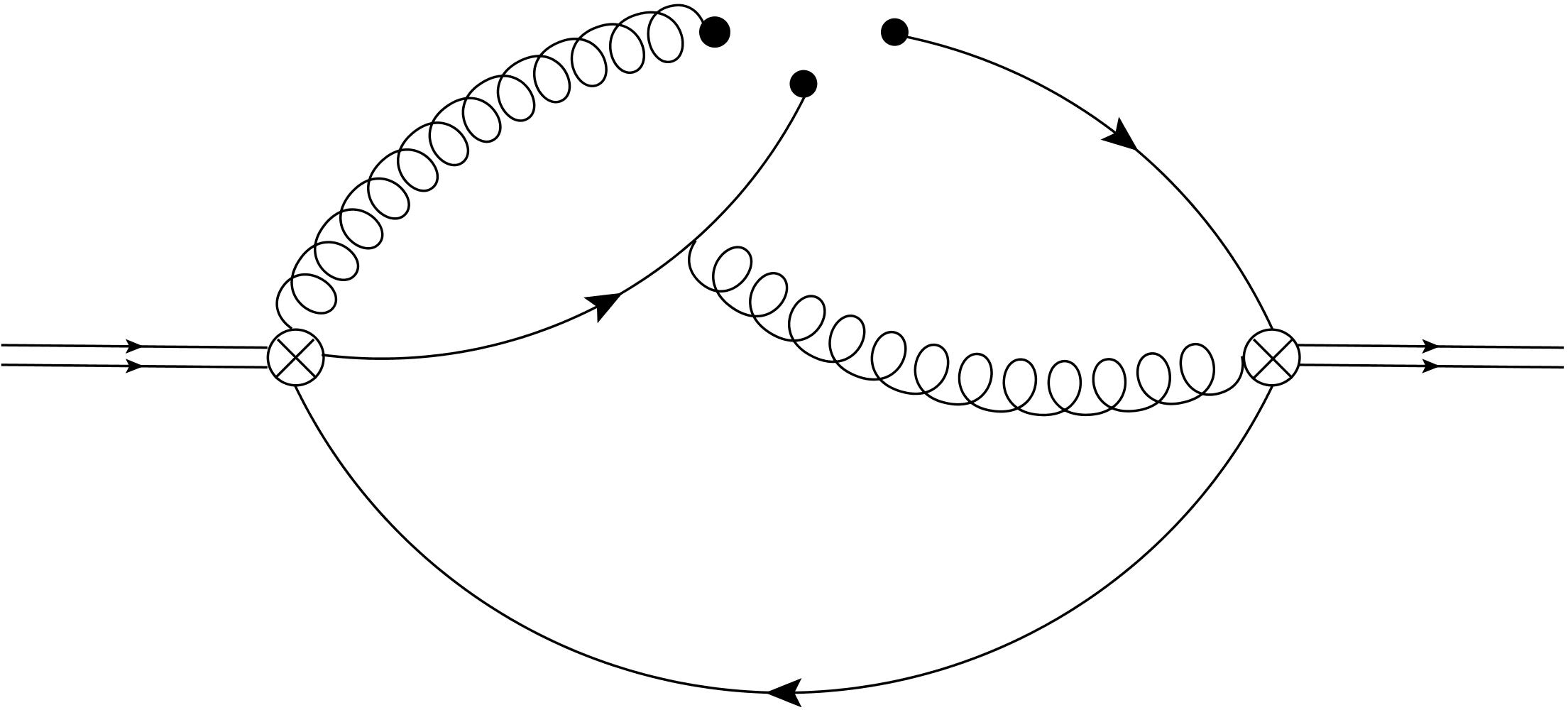}}\hfill
  \subfloat[Diagram XI (dimension-five)]{%
    \includegraphics[width=.26\textwidth]{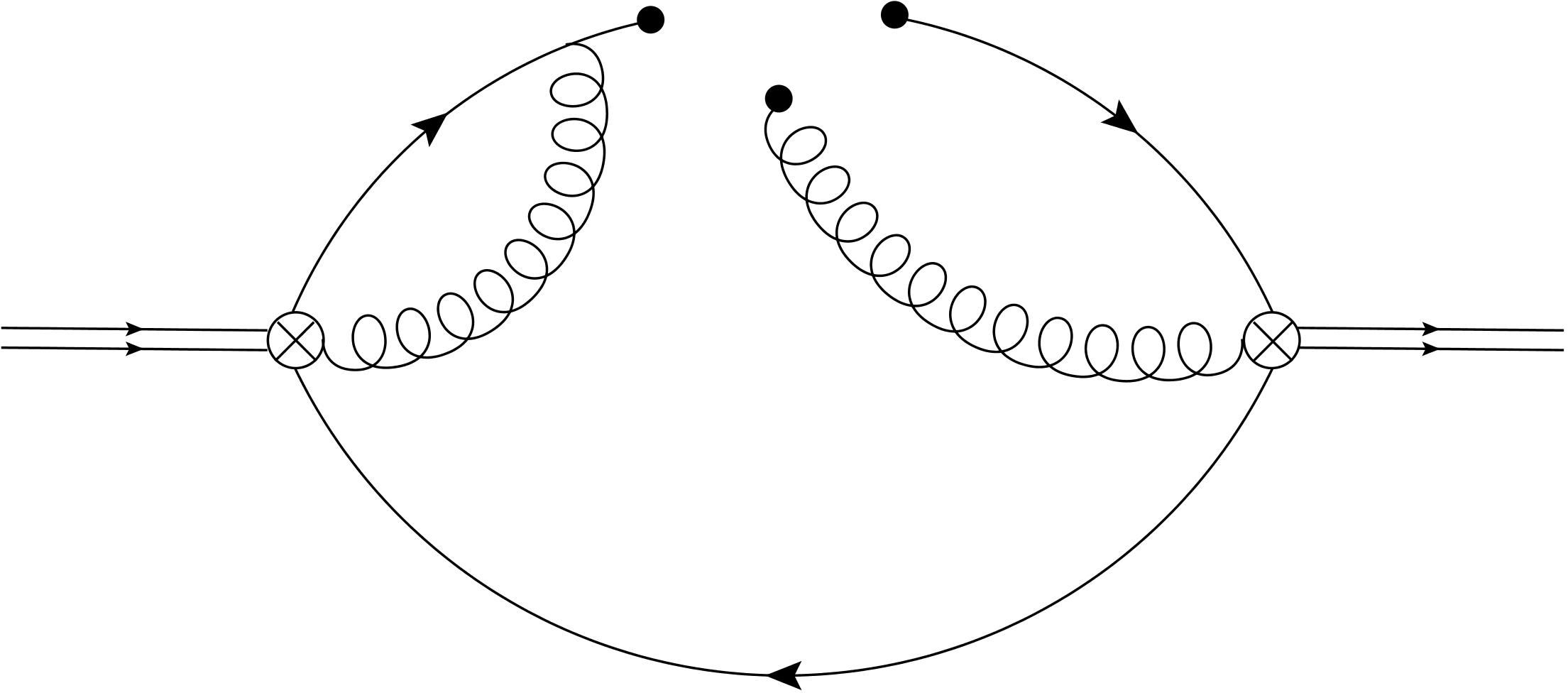}}\hfill
  \subfloat[Diagram XII (dimension-five)]{%
    \includegraphics[width=.28\textwidth]{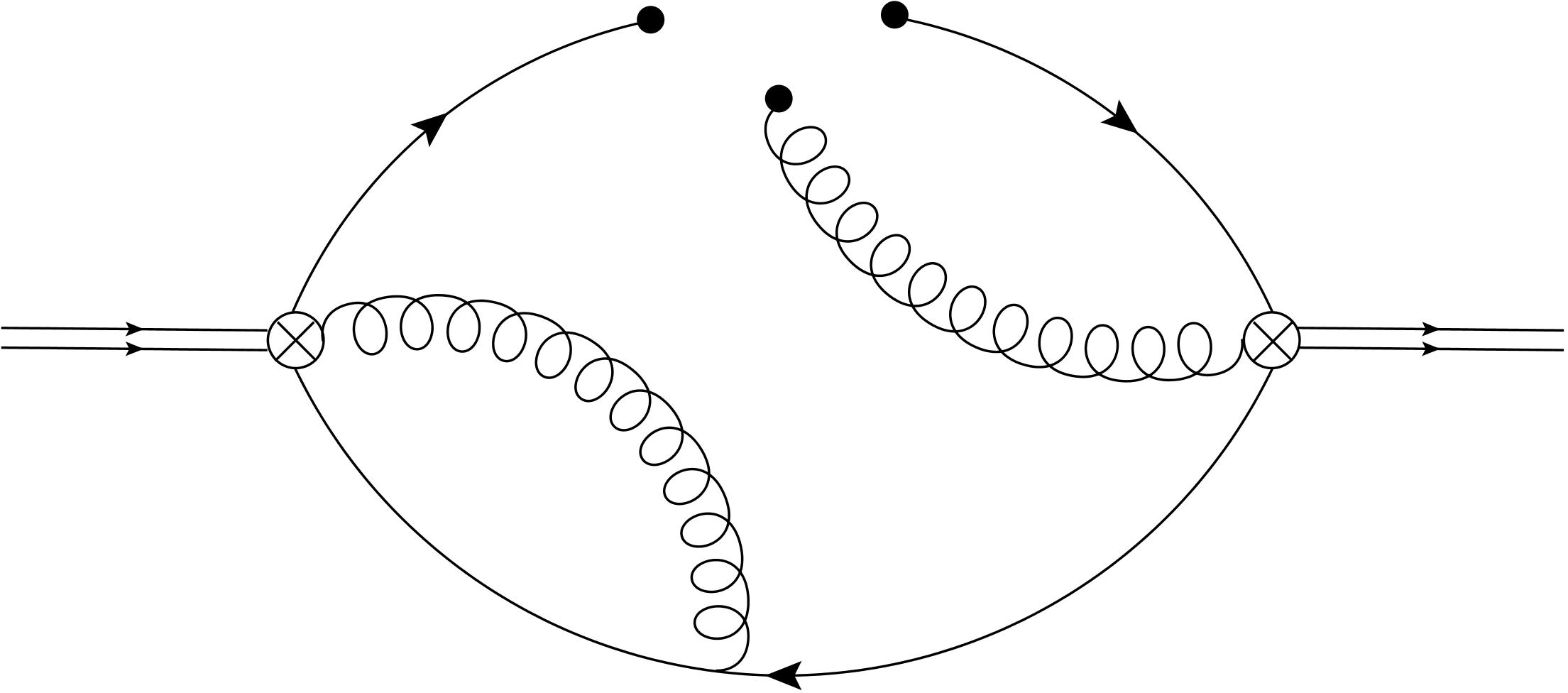}}\\
  \subfloat[Diagram XIII (dimension-five)]{%
    \includegraphics[width=.28\textwidth]{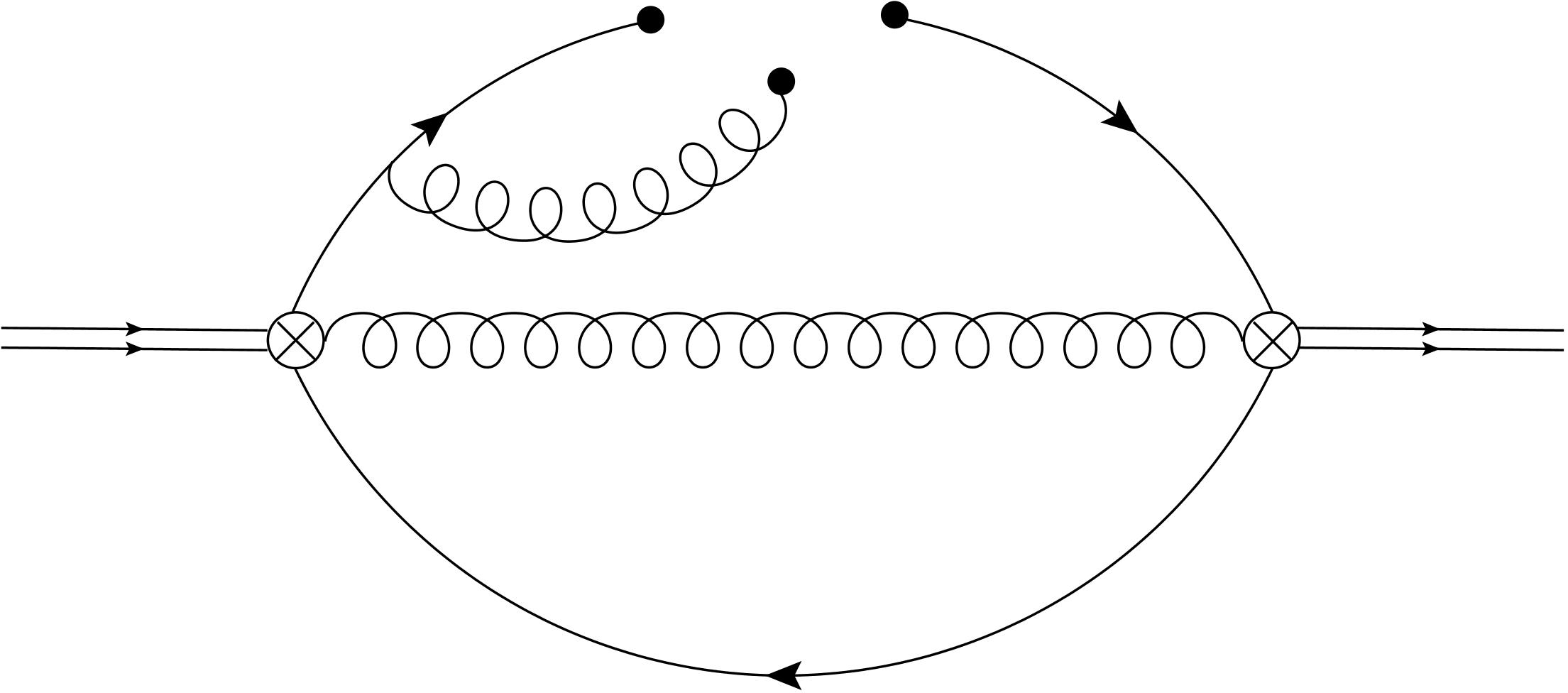}}\hfill
  \subfloat[Diagram XIV (dimension-five)]{%
    \includegraphics[width=.28\textwidth]{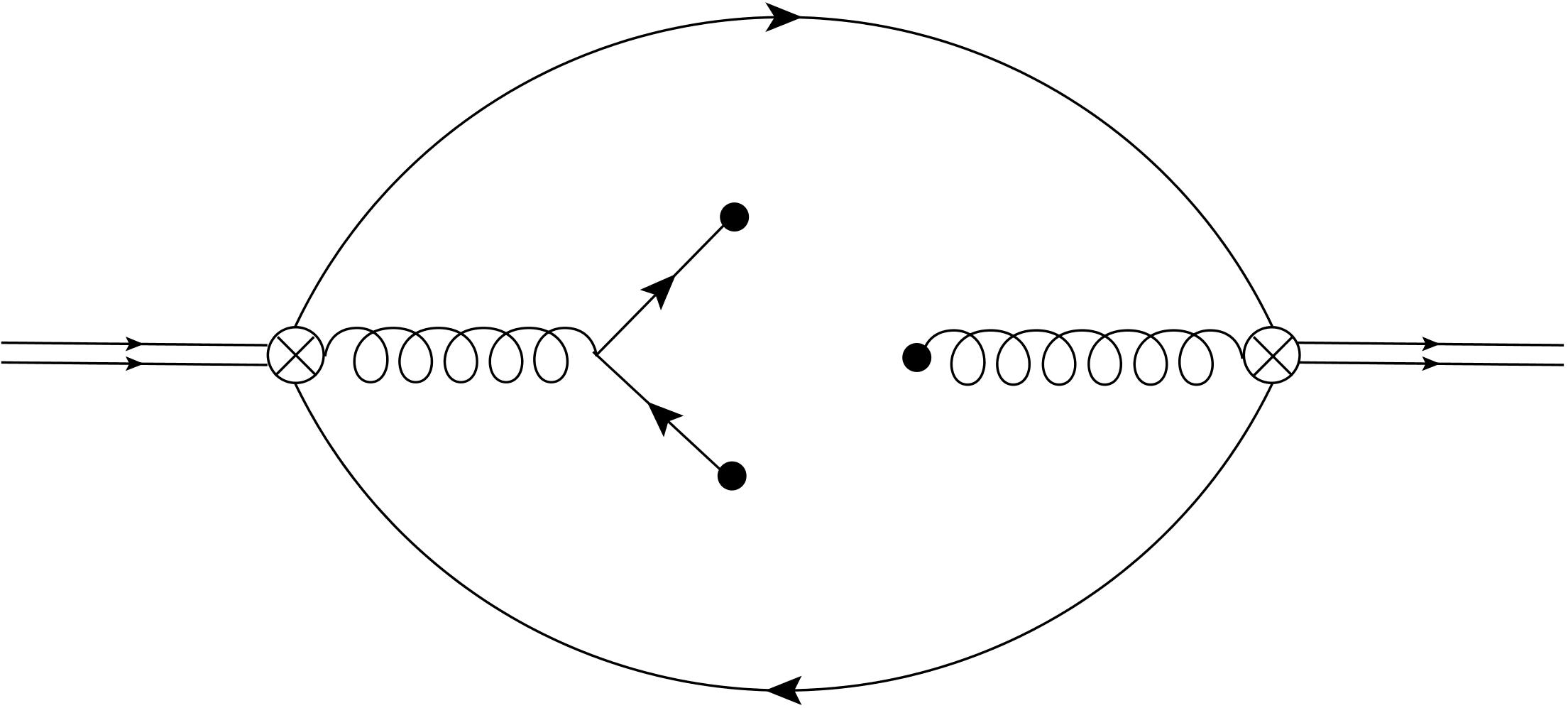}}\hfill
\caption{The Feynman diagrams calculated for the correlator~(\ref{correlator}). Feynman diagrams were created using
JaxoDraw~\cite{Binosi2004}.}
  \label{fig.feynman}
\end{figure*}

The calculation of~(\ref{correlator}) is performed in the framework of the 
operator product expansion (OPE),
\begin{equation}\label{opeFirst}
\vevomega{\tau\,\{ \mathcal{O}(x)\mathcal{O}(0) \}} = \sum_{n} C_{n}(x) 
\vevomega{:\!\mathcal{O}_{n}(0)\!:}. 
\end{equation}
In~(\ref{opeFirst}), the vacuum expectation value (VEV) of
a time-ordered, non-local product of composite operators is expanded in
a series, each term of which is a product of a perturbative Wilson coefficient $C_{n}(x)$ 
and a nonzero VEV of a local composite operator $\mathcal{O}_{n}(0)$, i.e.,
a condensate. 
The condensates parameterize the nonperturbative nature of the QCD vacuum,
and we include in our correlator calculations the following set:
\begin{widetext}
\begin{gather}
  \quarkthreed=\angled{\overline{q}_i^{\alpha} q_i^{\alpha}}
    \label{condensateQuarkThree}\\
  \gluonfourd=\angled{\alpha_s G^a_{\mu\nu} G^a_{\mu\nu}}\label{condensateGluonFour}\\
  \mixed=\angled{g_s \overline{q}_i^{\alpha}\sigma^{\mu\nu}_{ij}
    \lambda^a_{\alpha\beta} G^a_{\mu\nu} q_j^{\beta}}
    \label{condensateMixed}\\
  \gluonsixd=\angled{g_s^3 f^{abc} G^a_{\mu\nu}G^b_{\nu\rho}G^c_{\rho\mu}}\label{condensate_gluon_six}\\
  \quarksixd = \angled{\overline{q}_i^{\alpha} q_i^{\alpha}\overline{q}_j^{\beta} q_j^{\beta}},\label{condensateQuarkSix}
\end{gather}
\end{widetext}
respectively 
the 3d~quark condensate,
the 4d~gluon condensate,
the 5d~mixed condensate, 
the 6d~gluon condensate, and
the 6d~quark condensate.
In~(\ref{condensateQuarkThree})--(\ref{condensateQuarkSix}), 
superscripts on quark fields are colour  indices whereas subscripts are Dirac indices 
and $\sigma^{\mu\nu}=\frac{i}{2}[\gamma^{\mu},\gamma^{\nu}]$. 
Regarding Wilson coefficients, we consider LO calculations in $\alpha_s$,
and we compute $\mathcal{O}\left(m^2\right)$ light quark mass corrections 
to perturbation theory as a way to distinguish between the 
nonstrange- and strange-flavored cases, similar to~\cite{Ho2017}. 
Also, the values of~(\ref{condensateQuarkThree}), 
(\ref{condensateMixed}), and~(\ref{condensateQuarkSix}) depend on whether
the light quarks are nonstrange or strange.
The diagrams representative of the correlation function calculation are displayed 
in Figure~\ref{fig.feynman}.
We use dimensional regularization in $D=4+2\epsilon$ dimensions at 
$\overline{\mathrm{MS}}$ renormalization scale $\mu$.
The program TARCER~\cite{Mertig1998} is utilized to reduce the resulting integrals to a selection of 
well-known master integrals using the Tarasov recurrence relations~\cite{Tarasov1996, Tarasov1997}.

All of the correlators defined between~\eqref{hybridcurrent} 
and Table~\ref{JPC_table} can be written in general as
\begin{widetext}
\begin{equation}\label{generalcorrelator}
\begin{split}
\Pi\left(q^2\right) = & \alpha_{s} \left( A_{1} q^6 + A_{2} m^{2} q^{4}\right)\left\{\mathrm{log}\left(\frac{-q^2}{\mu^2}\right) + \frac{1}{2\epsilon}\right\} \\
& + \Big(A_{4} q^{2} \gluonfourd + \alpha_{s}(A_{3} q^{2} m \quarkthreed + A_{5} \quarkthreed^{2} + A_{6} \gluonsixd + A_{7} m \mixed)\Big)\\
& \qquad \times\left\{\mathrm{log}\left(\frac{-q^2}{\mu^2}\right) + \frac{1}{\epsilon}\right\} \\
& +\alpha_{s}\left( B_{1} q^{6} + B_{2} m^{2} q^{4} + B_{3} q^{2} m \quarkthreed + B_{4} q^{2} \gluonfourd \right. \\
& \qquad \qquad + \left. B_{5} \quarkthreed^{2} + B_{6} \gluonsixd + B_{7} m \mixed\right)
\end{split}
\end{equation}
\end{widetext}
where we have suppressed the superscript $(J)$ on the left-hand side.
The coefficients $A_{i}$ and $B_{j}$ contained in~\eqref{generalcorrelator} 
are given in Tables \ref{A_coefficients} and \ref{B_coefficients} respectively.
We note that, as Diagram~IV has no loops, $A_5$ is trivially zero.
In all channels, perturbation theory, the 3d quark condensate term, and the
4d gluon condensate term were benchmarked against~\cite{Govaerts1985}.
The $0^{--}$ and $1^{-+}$ correlators were benchmarked against~\cite{Latorre1987}.

\renewcommand{\arraystretch}{1.4}
\begin{table*}[htbp]
\centering
\caption{Coefficients of the logarithmic and divergent terms of the perturbative and condensate contributions to the correlation function~(\ref{generalcorrelator})
for the $J^{PC}$ summarized in Table~\ref{JPC_table}.}
\label{A_coefficients}
\begin{tabular}{|c||c|c|c|c|c|c|c|c|}
\hline
 & $0^{++}$ & $1^{-+}$ & $0^{--}$ & $1^{+-}$ & $0^{-+}$ & $1^{++}$ & $0^{+-}$ & $1^{--}$ \\ \hline \hline
 $A_{1}$ & $-\frac{1}{480\pi^{3}}$ & $-\frac{1}{240\pi^{3}}$ & $-\frac{1}{480\pi^{3}}$ & $-\frac{1}{240\pi^{3}}$ & $-\frac{1}{480\pi^{3}}$ & $-\frac{1}{240\pi^{3}}$ & $-\frac{1}{480\pi^{3}}$ & $-\frac{1}{240\pi^{3}}$ \\ \hline
 $A_{2}$& 0 & $\frac{1}{12\pi^{3}}$ & $\frac{1}{16\pi^{3}}$ & $\frac{5}{48\pi^{3}}$ & 0 & $\frac{1}{12\pi^{3}}$ & $\frac{1}{16\pi^{3}}$ & $\frac{5}{48\pi^{3}}$ \\ \hline
 $A_{3}$& $\frac{1}{3\pi}$ & $-\frac{2}{9\pi}$ & $-\frac{1}{3\pi}$ & $-\frac{4}{9\pi}$ & $\frac{1}{3\pi}$ & $-\frac{2}{9\pi}$ & $-\frac{1}{3\pi}$ & $-\frac{4}{9\pi}$ \\ \hline
 $A_{4}$& $\frac{1}{24\pi}$ & $-\frac{1}{36\pi}$ & $\frac{1}{24\pi}$ & $-\frac{1}{36\pi}$ & $-\frac{1}{24\pi}$ & $\frac{1}{36\pi}$ & $-\frac{1}{24\pi}$ & $\frac{1}{36\pi}$ \\ \hline
 $A_{5}$& 0 & 0 & 0 & 0 & 0 & 0 & 0 & 0 \\ \hline
 $A_{6}$& 0 & 0 & 0 & 0 & 0 & 0 & 0 & 0 \\ \hline
 $A_{7}$& $\frac{1}{9\pi}$ & 0 & $\frac{11}{72\pi}$ & $-\frac{19}{72\pi}$ & $-\frac{1}{9\pi}$ & 0 & $-\frac{11}{72\pi}$ & $\frac{19}{72\pi}$ \\ \hline
\end{tabular}
\end{table*}

\begin{table*}[htbp]
\centering
\caption{Coefficients of the finite terms of the perturbative and condensate contributions to the correlation function~(\ref{generalcorrelator}) for the $J^{PC}$ summarized in Table~\ref{JPC_table}.}
\label{B_coefficients}
\begin{tabular}{|c||c|c|c|c|c|c|c|c|}
\hline
 & $0^{++}$ & $1^{-+}$ & $0^{--}$ & $1^{+-}$ & $0^{-+}$ & $1^{++}$ & $0^{+-}$ & $1^{--}$  \\ \hline \hline
$B_{1}$ & $\frac{97}{19200\pi^{3}}$ & $\frac{39}{3200\pi^{3}}$ & $\frac{97}{19200\pi^{3}}$ & $\frac{39}{3200\pi^{3}}$ & $\frac{19}{6400\pi^{3}}$ & $\frac{77}{9600\pi^{3}}$ & $\frac{19}{6400\pi^{3}}$ & $\frac{77}{9600\pi^{3}}$ \\ \hline
$B_{2}$ & $\frac{1}{32\pi^{3}}$ &  $-\frac{7}{32\pi^{3}}$ & $-\frac{55}{384\pi^{3}}$ & $-\frac{109}{384\pi^{3}}$ & $\frac{1}{32\pi^{3}}$ & $-\frac{13}{96\pi^{3}}$ & $-\frac{31}{384\pi^{3}}$ & $-\frac{23}{128\pi^{3}}$\\ \hline
$B_{3}$ & $-\frac{1}{2\pi}$ & $\frac{7}{27\pi}$ & $\frac{1}{6\pi}$ & $\frac{17}{27\pi}$ & $\frac{1}{6\pi}$ & $-\frac{5}{27\pi}$ & $-\frac{1}{2\pi}$ &  $-\frac{7}{27\pi}$\\ \hline
$B_{4}$ & $-\frac{13}{144\pi}$ & $\frac{11}{216\pi}$ & $-\frac{13}{144\pi}$ & $\frac{11}{216\pi}$ & $-\frac{5}{144\pi}$ & $\frac{7}{216\pi}$ & $-\frac{5}{144\pi}$ &  $\frac{7}{216\pi}$ \\ \hline
$B_{5}$ & $-\frac{4\pi}{3}$ & $\frac{4\pi}{9}$ & $\frac{4\pi}{3}$ & $-\frac{4\pi}{9}$ & 0 & $-\frac{8\pi}{9}$ & 0 & $-\frac{8\pi}{9}$ \\ \hline
$B_{6}$ & $-\frac{1}{192\pi^{2}}$ & $\frac{1}{192\pi^{2}}$ & $-\frac{1}{192\pi^{2}}$ & $\frac{1}{192\pi^{2}}$ & $\frac{5}{192\pi^{2}}$ & $-\frac{5}{192\pi^{2}}$ & $\frac{5}{192\pi^{2}}$ & $-\frac{5}{192\pi^{2}}$ \\ \hline
$B_{7}$ & $-\frac{461}{1728\pi}$ & $-\frac{83}{1728\pi}$ & $-\frac{731}{1728\pi}$ & $\frac{1019}{1728\pi}$ & $-\frac{217}{1728\pi}$ & $\frac{265}{1728\pi}$ & $\frac{41}{1728\pi}$ & $\frac{71}{1728\pi}$ \\ \hline
\end{tabular}
\end{table*}

\section{QCD Sum-Rules} \label{III}
%
Each function $\Pi^{(J)}(q^2)$ defined in~(\ref{spin_breakdown}) 
satisfies a dispersion relation at Euclidean momentum $Q^2=-q^2>0$,
\begin{equation}\label{dispersionRelation}
  \Pi\left(Q^2\right) = Q^8\int_{t_0}^{\infty}\! 
  \frac{\frac{1}{\pi}\Im\Pi(t)}{t^4\left(t+Q^{2}\right)}\,\mathrm{d}t + \cdots,
\end{equation}
where we have again suppressed the superscript $(J)$.
In~(\ref{dispersionRelation}), $t_0$ is a hadron production threshold
and $\cdots$ are subtraction constants, together a third degree polynomial in $Q^2$.
Equation~(\ref{dispersionRelation}) connects theoretical predictions of QCD,
i.e., $\Pi(Q^2)$ on the left-hand side, to properties of hadrons contained in $\Im\Pi(t)$, 
the hadronic spectral function, on the right-hand side.

Regarding~(\ref{dispersionRelation}),
to eliminate subtraction constants and to accentuate the low-energy region
of the integral on the right-hand side, some transformation is typically applied.
A popular choice is to formulate unsubtracted LSRs of (usually nonnegative) 
integer weight $k$,
\begin{widetext}
\begin{equation}\label{unsubtractedLSR}
  \lsr_k(M_B) = M_B^{2}\lim_{\stackrel{N,Q^2\rightarrow\infty}{M_B^{2}=Q^2/N}}
  \frac{\big(-Q^2\big)^N}{\Gamma(N)}\bigg(\frac{d}{dQ^2}\bigg)^N\Big\{(-Q^2)^k \Pi(Q^2)\Big\},
\end{equation}
\end{widetext}
at Borel parameter $M_B$~\cite{Shifman:1978bx,Shifman:1978by,Reinders:1984sr,Narison2004}.
Details on how to evaluate~(\ref{unsubtractedLSR}) for a correlator such as~(\ref{generalcorrelator}),
denoted $\Pi^{\text{QCD}}$ from here on to emphasize that it is a quantity calculated using QCD, 
can be found in the literature (e.g.,~\cite{Shifman:1978bx}).
The result is
\begin{equation}\label{lsrIntegral}
  \lsr_k(M_B) = \int_0^{\infty}\! t^k 
   e^{-t/M_B^2}
  \frac{1}{\pi}\Im\Pi^{\text{QCD}}(t)\,\dt
\end{equation}
for $k\in\{0,\,1,\,2,\ldots\}$ and where

\begin{equation}
\begin{split}\label{ImPart}
  \frac{1}{\pi}\Im\Pi^{\text{QCD}}(t) =   
    - & A_1 \as t^3 - A_2 \as m^2 t^2 \\
    & - A_3 \as t \mquarkthreed 
     -A_4 t \gluonfourd \\ & - A_7 \as m \mixed.
\end{split}
\end{equation}
Recall, the $A_i$ are given in Table~\ref{A_coefficients}. 

In~(\ref{dispersionRelation}), we impose on $\Im\Pi(t)$ a general resonances-plus-continuum model 
with onset of the QCD continuum at threshold $s_0$,
\begin{equation}\label{general_model}
\Im\Pi(t) = \rho^\mathrm{had}(t) + \theta\left(t-s_0\right)\mathrm{Im}\Pi^\mathrm{QCD}(t),
\end{equation}
where $\rho^{\text{had}}(t)$ represents the resonance content of the hadronic spectral function  
and $\theta(t)$ is the Heaviside step function.
To isolate the resonance contributions to the LSRs, we consider (continuum-) subtracted LSRs
\begin{equation}\label{subtractedLSRdefn}
  \lsr_k(M_B,\,s_0) = \lsr_k(M_B) 
    - \int_{s_0}^{\infty}\! t^k 
 e^{-t/M_B^2}
 \frac{1}{\pi}\Im\Pi^{\text{QCD}}(t)\,\dt.
\end{equation}
Then, Equations~(\ref{dispersionRelation})--(\ref{lsrIntegral}),
(\ref{general_model}), and~(\ref{subtractedLSRdefn}) together imply that 
\begin{equation}\label{masterFormulaLSR}
  \lsr_k(M_B,\,s_0) = \int_{t_0}^{\infty}\! t^k 
 e^{-t/M_B^2}
  \frac{1}{\pi}\rho^{\text{had}}(t)\,\dt
\end{equation}
where 
\begin{equation}\label{subtractedLSR}
   \lsr_k(M_B,\,s_0) = \int_{0}^{s_0}\! t^k 
  e^{-t/M_B^2} 
  \frac{1}{\pi}\Im\Pi^{\text{QCD}}(t)\,\dt
\end{equation}
and (again) $\Im\Pi^{\text{QCD}}(t)$ is given in~(\ref{ImPart}).

There are a number of interesting observations we can make concerning the LSRs
of light quarkonium hybrids.  In particular,
the 6d gluon condensate terms do not contain a logarithm, i.e., $A_6=0$ for all 
$J^{PC}$ values considered (see Table~\ref{A_coefficients}),
and hence do not contribute to the imaginary part~(\ref{ImPart}).  
This result is surprising: both Diagrams~V and~VI (see Figure~\ref{fig.feynman})
have logarithmic contributions, but they cancel when the two diagrams are added together.
Thus, the LO 6d gluon condensate terms cannot stabilize
light quarkonium hybrid LSRs analyses as they have done in some heavy quarkonium hybrid 
analyses~\cite{Qiao2012,Berg2012,Chen:2013zia,Chen:2013eha}.

Another observation relates to the mixed condensate contributions.
Using~(\ref{unsubtractedLSR}), if we try to formulate $k=-1$ (i.e., lower-weight) unsubtracted LSRs, 
we get a piece that formally looks like the right-hand side of~(\ref{lsrIntegral}) at $k=-1$
and another piece:
\begin{equation}\label{extraTerms}
  - B_5 \quarkthreed^2 - B_6 \gluonsixd 
  - \left( \frac{A_7}{\epsilon} + B_7 \right) m \mixed.
\end{equation}
If $A_7\neq 0$, then neither piece is well-defined: the integral from~(\ref{lsrIntegral})
diverges and~(\ref{extraTerms}) contains a $\epsilon^{-1}$ field theory divergence.
But for $J^{PC}\in\{1^{-+},\,1^{++}\}$, we find that $A_7=0$ 
which allows for the construction of lower-weight LSRs
in these two channels.
Unlike the $k=0$ LSRs, the $k=-1$ LSRs do receive contributions from the
6d quark and gluon condensates as both $B_5$ and $B_6$ are nonzero. 
An analysis of these $k=-1$ LSRs does require some knowledge
of the subtraction constants in~(\ref{dispersionRelation}).

As noted in Section~\ref{I}, in the multi-channel LSRs analysis of~\cite{Govaerts1985},
the $0^{+-}$, $0^{-+}$, $1^{+-}$, and $1^{-+}$ sectors were unstable.
The $1^{-+}$ has since been stabilized using lower-weight LSRs~\cite{Latorre1987}, and
the $0^{-+}$ has been stabilized~\cite{Huang1999}
using a different current than that used in~\cite{Govaerts1985}. 
That leaves the non-exotic $1^{+-}$ and the exotic $0^{+-}$ channels.  
Given the GlueX emphasis on exotics
and the possible complicated features of mixing between hybrids and conventional quark mesons in the $1^{+-}$ channel, we focus our attention on $0^{+-}$ light quarkonium hybrids.
Attempts to stabilize the $0^{+-}$ channel have involved higher-dimension currents combined with lower-weight sum-rules requiring estimation of the dispersion-relation low-energy constant 
within the analysis~\cite{Braun1986}.  
Because higher-dimension currents tend to enhance the continuum, the mass determination combined with an estimated low-energy term merits further study.

As in~\cite{Govaerts1985}, we perform a conventional single narrow 
resonance (SNR) LSRs analysis of the $0^{+-}$ channel by letting
\begin{equation}\label{singleNarrowResonance}
  \rho^{\text{had}}(t) = \pi f^2 \delta(t-m_H^2)
\end{equation}
in~(\ref{masterFormulaLSR})
where $f$ is the resonance coupling and $m_H$ is its mass.
We include our higher-dimension condensate contributions as well as updated QCD parameter values, 
yet the analysis remains unstable. 
The 5d mixed condensate term in the LSRs is small, and, as noted above, the 6d condensates
do not contribute at all.
In~\cite{Govaerts1985}, it was suggested that the instability in this channel could be 
related to a distribution of resonance strength. 
To investigate this possibility, we use GSRs, an alternative to LSRs 
which provide a fundamentally different weighting of the hadronic spectral function
that makes them well-suited to analyzing distributed resonance strength hadron models.
Unsubtracted GSRs of integer weight $k$ are defined as~\cite{Bertlmann1985}
\begin{widetext}
\begin{multline}\label{unsubtractedGSR}
  G_k(\hat{s},\,\tau) = \sqrt{\frac{\tau}{\pi}}
  \lim_{\stackrel{N,\Delta^2\rightarrow\infty}{\tau=\Delta^2/(4N)}}
  \frac{\big(-\Delta^2\big)^N}{\Gamma(N)}\\
  \times\bigg(\frac{d}{d\Delta^2}\bigg)^N
  \left\{\frac{(\hat{s}+i\Delta)^k \Pi(-\hat{s}-i\Delta)-(\hat{s}-i\Delta)^k \Pi(-\hat{s}+i\Delta)}%
  {i\Delta}\right\}.
\end{multline}
\end{widetext}
Details on how to evaluate~(\ref{unsubtractedGSR}) for a correlator such as~(\ref{generalcorrelator})
can be found in~\cite{Bertlmann1985,Orlandini2001,Harnett2001}.
The result is
\begin{equation}\label{gsrIntegral}
  G_k(\hat{s},\,\tau) = \frac{1}{\sqrt{4\pi\tau}}\int_0^{\infty}\! t^k 
  e^{-\frac{(\hat{s}-t)^2}{4\tau}} \frac{1}{\pi}\Im\Pi^{\text{QCD}}(t)\,\dt
\end{equation}
for $k\in\{0,\,1,\,2,\ldots\}$ and where $\frac{1}{\pi}\Im\Pi^{\text{QCD}}(t)$ is given in~(\ref{ImPart}).
Subtracted GSRs are defined in much the same way as subtracted LSRs leading to
the following GSRs analogues of~(\ref{masterFormulaLSR}) and (\ref{subtractedLSR}):
\begin{equation}\label{masterFormulaGSR}
  G_k(\hat{s},\,\tau,\,s_0) = \frac{1}{\sqrt{4\pi\tau}}\int_{t_0}^{\infty}\! t^k 
  e^{-\frac{(\hat{s}-t)^2}{4\tau}} \frac{1}{\pi}\rho^{\text{had}}(t)\,\dt
\end{equation}
where
\begin{equation}\label{subtractedGSR}
\begin{split}
  G_k(\hat{s},\,\tau,\,s_0) = \frac{1}{\sqrt{4\pi\tau}}\int_0^{\infty}\! & t^k 
  e^{-\frac{(\hat{s}-t)^2}{4\tau}} \\
  &\times \frac{1}{\pi}\Im\Pi^{\text{QCD}}(t)\,\dt.
  \end{split}
\end{equation}

The difference between (\ref{masterFormulaLSR})--(\ref{subtractedLSR}) 
and (\ref{masterFormulaGSR})--(\ref{subtractedGSR}) 
is in the kernel of the integrals: a decaying exponential for LSRs and a Gaussian for GSRs. 
The two sum-rules represent fundamentally different weightings of the spectral function $\rho^{\text{had}}(t)$; whereas in the LSRs have a duality interval  of width $\sim 1/M_B^2$ near the low-energy threshold of the spectral function, the GSRs have a duality interval of width $\sim \sqrt{2\tau}$ near $\hat s$ (\ref{masterFormulaGSR}).
In the $\tau \rightarrow 0^+$ limit, we have
\begin{equation}
  \lim_{\tau \rightarrow 0^+} \frac{1}{\sqrt{4\pi\tau}} 
  e^{-\frac{(\hat{s}-t)^2}{4\tau}} = \delta\left(\hat{s}-t\right), 
\end{equation}
which, when applied to~(\ref{masterFormulaGSR}), yields
\begin{equation}
  \lim_{\tau\rightarrow 0^+} G_k(\hat{s},\,\tau,\,s_0)
  = \hat{s}^k \frac{1}{\pi}\rho^{\text{had}}(\hat{s})\ \text{for}\ \hat{s}>t_0.
\end{equation}
Hence, at least in principle, $\rho^{\text{had}}(t)$ can be extracted directly from GSRs.
Realistically, the $\tau\rightarrow 0^+$ limit cannot be achieved, however, 
because, through renormalization-group (RG) improvement (see Section~\ref{V}), 
the renormalization scale at which $\alpha_s$ is evaluated
decreases with decreasing $\tau$~\cite{Bertlmann1985}.
Nevertheless, it is desirable to use low values of $\tau$ 
to minimize the smearing of $\rho^{\text{had}}(t)$ by the kernel of the GSRs.
To further emphasize this, 
we draw upon an analogy introduced in the seminal GSRs paper~\cite{Bertlmann1985}. 
Gaussian sum-rules satisfy the classical heat equation
\begin{equation}\label{heat}
  \frac{\partial^2 G_k\left(\hat{s},\,\tau,\,s_0\right)}{\partial \hat{s}^2} 
  = \frac{\partial G_k\left(\hat{s},\,\tau,\,s_0\right)}{\partial \tau},
\end{equation}
reinterpreting 
the parameter $\hat{s}$ as ``position'', 
the Gaussian width $\tau$ as ``time'',
and the GSRs $G_k(\hat{s},\,\tau,\,s_0)$ as ``temperature''.
The smaller the value of $\tau$ (i.e. the less ``time'' that has passed), 
the better we may assess the original (i.e., $\tau\rightarrow 0^+$) 
``temperature'' distribution (i.e., $\hat{s}^k\frac{1}{\pi}\rho^{\text{had}}(\hat{s})$).

Compared to LSRs, GSRs permit greater access to the structure of $\rho^{\text{had}}(t)$.
The LSRs methodology is specifically formulated to accentuate the ground state
region of the hadronic spectral function while suppressing  higher energies.
With GSRs, this need not be the case as $\hat{s}$, the position of the
Gaussian kernel's peak, is a free parameter.  
By varying $\hat{s}$, GSRs can probe 
a wide region of the hadronic spectral function
with the same sensitivity as the ground state region.
As such, GSRs are generally preferable to LSRs when studying distributed resonance strength models, as demonstrated in the successful analysis of the $\rho$ meson using GSRs methodology~\cite{Orlandini2001}.
Integrating~(\ref{masterFormulaGSR}) with respect to $\hat{s}$ gives
\begin{equation}\label{fesrConstraint}
  \int_{-\infty}^{\infty}\! G_k(\hat{s},\,\tau,\,s_0)\,\mathrm{d}\hat{s}
  = \int_{t_0}^{\infty}\! t^k \frac{1}{\pi} \rho^{\text{had}}(t)\,\dt
\end{equation}
from which we recognize the quantity on the left-hand side as the finite-energy
sum-rule of weight $k$.
As shown in~\cite{Bertlmann1985}, a resonance plus continuum model evolved through the diffusion equation only reproduces the QCD prediction at large energy scales if
$s_0$ is constrained by~(\ref{fesrConstraint}).
To isolate the information contained in the GSRs formalism that is independent 
of~(\ref{fesrConstraint}),
we consider normalized Gaussian sum-rules (NGSRs)~\cite{Orlandini2001}
\begin{equation}\label{gsrNormalized}
  N_{k}\left( \hat{s},\tau,s_0\right) 
  = \frac{G_{k}\left(\hat{s},\tau,s_0\right)}{M_{k,0}\left(\tau,s_0\right)},
\end{equation}
i.e., GSRs scaled by their $0^{\text{th}}$-order moments $M_{k,0}(\hat{s},\,\tau)$ where, in general,
\begin{equation}\label{gsrMoment}
M_{k,n}\left(\tau,s_0\right) = \int_{-\infty}^{\infty}\!\hat{s}^{n} G_{k}\left(\hat{s},\tau,s_0\right)\, \mathrm{d}\hat{s}.
\end{equation}
Combining~(\ref{masterFormulaGSR}), (\ref{fesrConstraint}), and~(\ref{gsrNormalized}), 
we get a NGSRs analogue of~(\ref{masterFormulaGSR}),
\begin{equation}\label{masterFormulaNGSR}
  N_k(\hat{s},\,\tau,\,s_0) = \frac{\frac{1}{\sqrt{4\pi\tau}}\int_{t_0}^{\infty} t^k 
  e^{-\frac{(\hat{s}-t)^2}{4\tau}} \frac{1}{\pi}\rho^{\text{had}}(t)\,\dt}%
  {\int_{t_0}^{\infty} t^k \frac{1}{\pi}\rho^{\text{had}}(t)\,\dt}.
\end{equation}
Finally, to emphasize the low-energy region of the spectral function, 
we work with the lowest-weight sum-rules (i.e., $k=0$) as in previous 
applications of GSRs to the prediction of resonance properties~\cite{Orlandini2001,Harnett2001}.
\section{H\"older Inequality}\label{IV}
Previous investigations of hadronic systems using LSRs
have employed H\"older inequalities to restrict the 
set of allowed $\tau$ and $s_0$ values~\cite{Wang2017, Shi2000,Benmerrouche1996}. 
The H\"older Inequality is expressed generally as
\begin{equation}\label{holder_general}
\begin{split}
\abs{\int_{t_1}^{t_2}\! f\left(t\right)g\left(t\right)\,\mathrm{d}\mu} \leq  &
\left( \int_{t_1}^{t_2}\!\abs{f\left(t\right)}^{p}\,\mathrm{d}\mu \right)^{\frac{1}{p}}\\
&\times \left( \int_{t_1}^{t_2}\!\abs{g\left(t\right)}^{q}\,\mathrm{d}\mu \right)^{\frac{1}{q}}
\end{split}
\end{equation}
under the condition
\begin{equation}
\frac{1}{p} + \frac{1}{q} = 1
\end{equation}
and where $d\mu$ is an arbitrary integration measure.
From positivity of the hadronic spectral function for diagonal correlators,
we can use $\Im\Pi^{\text{QCD}}(t)>0$ to form an integration measure.
Substituting this integration measure into~(\ref{holder_general}) leads to 
restrictions on the allowed values of $\hat{s}$, $\tau$, and $s_0$ in the GSRs. 
We consider the inequality~(\ref{holder_general}) with the assignments
\begin{gather}
\mathrm{d}\mu = \Im\Pi^{\text{QCD}}(t)\mathrm{d}t\\
f\left(t\right) = t^{\alpha} \left(\frac{e^{-\frac{\left(\hat{s}-t\right)^2}{4\tau}}}{\sqrt{4\pi \tau}}\right)^{a}\\
g\left(t\right) = t^{\beta} \left(\frac{e^{-\frac{\left(\hat{s}-t\right)^2}{4\tau}}}{\sqrt{4\pi \tau}}\right)^{b}\\
t_{1} = t_{0}, \, t_{2}  = s_0\\
a + b = 1 \label{holder_constraint}
\end{gather}
where $\alpha+\beta$ is a non-negative integer.
Defining 
\begin{equation}\label{tauDefns}
\tau_{1}=\frac{\tau}{ap}\ \text{and}\ \tau_{2}=\frac{\tau}{bq}, 
\end{equation}
the inequality~\eqref{holder_general} becomes
\begin{equation}\label{holder_gsr}
\begin{split}
G_{\alpha+\beta}\left(\tau,\, \hat{s},\, s_{0}\right)
\leq &
\left(\frac{\tau_1}{\tau}\right)^{\frac{1}{2p}}\left(\frac{\tau_2}{\tau}\right)^{\frac{1}{2q}}\\&\times G_{\alpha p}^{\frac{1}{p}}\left(\tau_{1},\, \hat{s},\, s_{0}\right) G_{\beta q}^{\frac{1}{q}}\left(\tau_{2},\, \hat{s},\, s_{0}\right)
\end{split}
\end{equation}
where we have used $G_k\left(\tau,\,\hat{s},\,s_0\right)>0$, 
the weakest constraint on the GSRs that emerges from positivity of the spectral function.
We define $\omega$ as follows:
\begin{equation}\label{omegaDefn}
\omega=\frac{1}{p} \Longleftrightarrow 1-\omega=\frac{1}{q}, \ 0 < \omega <1
\end{equation}
and consider~(\ref{holder_gsr}) with zero-weight GSRs (i.e., $\alpha =\beta = 0$),
\begin{equation}\label{holder_gsr_zero}
\begin{split}
G_{0}\left(\tau,\,\hat{s},\,s_{0}\right)
\leq &
\left(\frac{\tau_1}{\tau}\right)^{\frac{\omega}{2}}\left(\frac{\tau_2}{\tau}\right)^{\frac{1-\omega}{2}}  \\
& \times G_{0}^{\omega}\left(\tau_{1}, \hat{s}, s_{0}\right) G_{0}^{1-\omega}\left(\tau_{2}, \hat{s}, s_{0}\right).
\end{split}
\end{equation}
Equations~(\ref{holder_constraint}), (\ref{tauDefns}), and~(\ref{omegaDefn}) 
together imply that
\begin{equation}
\tau = \frac{\tau_{1}\tau_{2}}{(1-\omega)\tau_{1}+\omega \tau_{2}}
\end{equation}
which, when substituted into~(\ref{holder_gsr_zero}), gives
\begin{widetext}
\begin{align}\label{holder_gsr_final}
G_{0}\left(\frac{\tau_{1}\tau_{2}}{(1-\omega)\tau_{1}+\omega \tau_{2}},\, \hat{s},\, s_{0}\right)
\leq &
\left(\frac{(1-\omega)\tau_{1}+\omega \tau_{2}}{\tau_{2}}\right)^{\frac{\omega}{2}}\left(\frac{(1-\omega)\tau_{1}+\omega \tau_{2}}{\tau_{1}}\right)^{\frac{1-\omega}{2}}\nonumber\\
& \times 
G_{0}^{\omega}\left(\tau_{1},\, \hat{s},\, s_{0}\right) 
G_{0}^{1-\omega}\left(\tau_{2},\, \hat{s},\, s_{0}\right).
\end{align}
\end{widetext}
Following~\cite{Benmerrouche1996}, we set
\begin{gather}
\tau_1 =  \tau\\
\tau_2 =  \tau + \delta \tau
\end{gather}
which implies
\begin{widetext}
\begin{multline}\label{holder_gsr_local}
0\leq G_{0}\left(\frac{\tau(\tau + \delta \tau)}{\omega (\tau + \delta \tau)+(1-\omega)\tau},\, \hat{s},\, s_{0}\right)
-\left(\frac{\omega (\tau + \delta \tau)+(1-\omega)\tau}{(\tau + \delta \tau)}\right)^{\frac{\omega}{2}}\left(\frac{\omega (\tau + \delta \tau)+(1-\omega)\tau}{\tau}\right)^{\frac{1-\omega}{2}}\\
\times G_{0}^{\omega}\left(\tau,\, \hat{s},\, s_{0}\right) G_{0}^{1-\omega}\left(\tau + \delta \tau,\, \hat{s},\, s_{0}\right).
\end{multline}
\end{widetext}

We can perform a local analysis of~(\ref{holder_gsr_local})
by expanding about $\delta \tau = 0$,

\begin{widetext}
\begin{multline}\label{holderIntermediate}
0\leq
\frac{\omega(\omega - 1)\left(1-2\tau^{2}\left(\frac{G^{'}_{0}(\hat{s},\,\tau,\, s_0)}{G_{0}(\hat{s},\,\tau,\, s_0)}\right)^{2}+2\tau \left(2\left(\frac{G^{'}_{0}(\hat{s},\,\tau,\, s_0)}{G_{0}(\hat{s},\,\tau,\, s_0)}\right)+\tau \left(\frac{G^{''}_{0}(\hat{s},\,\tau,\, s_0)}{G_{0}(\hat{s},\,\tau,\, s_0)}\right)\right)\right)(\delta \tau)^{2}}{4\tau^{2}} \\
+ \mathcal{O}\left((\delta \tau)^{3}\right),
\end{multline}
\end{widetext}
where primes indicate $\tau$-derivatives.
Then,~\eqref{omegaDefn} and~(\ref{holderIntermediate}) together imply
\begin{widetext}
\begin{equation}\label{holder_coefficient}
H\left(\hat s,\,\tau,\, s_0\right)\equiv 1-2\tau^{2}\left(\frac{G^{'}_{0}(\hat{s},\,\tau,\, s_0)}{G_{0}(\hat{s},\,\tau,\, s_0)}\right)^{2}+2\tau \left(2\left(\frac{G^{'}_{0}(\hat{s},\,\tau,\, s_0)}{G_{0}(\hat{s},\,\tau,\, s_0)}\right)+\tau \left(\frac{G^{''}_{0}(\hat{s},\,\tau,\, s_0)}{G_{0}(\hat{s},\,\tau,\, s_0)}\right)\right) \geq 0.
\end{equation}
\end{widetext}
At some $(\tau,\,\hat{s},\,s_0)$,
if the GSR $G_0(\hat{s},\,\tau,\,s_0)$ is to be consistent with a positive hadronic spectral function, 
then it must satisfy the inequality~\eqref{holder_coefficient}.

\section{Analysis Methodology and Results}\label{V}
Before we can analyze $0^{+-}$ light quarkonium hybrids using~(\ref{masterFormulaNGSR}),
we need to discuss the QCD parameters appearing in~(\ref{generalcorrelator}),
i.e., the coupling, the quark mass, and the QCD condensates.

To implement RG improvement 
we replace $\alpha_s$ and $m$ in~(\ref{generalcorrelator})
by one-loop, $\overline{\text{MS}}$ running quantities~\cite{Bertlmann1985}.
In our analysis, we use the QCD running coupling anchored at the $\tau$-lepton mass,
\begin{equation}\label{running_coupling}
\alpha_{s}(\mu) = \frac{\alpha_{s}\left(M_{\tau}\right)}{1+\frac{\alpha_{s}\left(M_{\tau}\right)}{12\pi}\left(33-2n_f\right) \log(\frac{\mu^2}{M_{\tau}^2})},
\end{equation}
where
we use PDG~\cite{PDG2017} values for the $\tau$ mass and 
\begin{equation}
  \alpha_{s}\left(M_{\tau}\right) = 0.325\pm 0.015.
\end{equation}
For the light quark masses, we use
\begin{equation}\label{running_light_quark}
  m(\mu)=m(2~\gev)\left(\frac{\as(\mu)}{\as(2~\gev)}\right)^{\frac{12}{33-2n_f}},
\end{equation}
where
\begin{equation}
\begin{split}
  m(2~\gev) &= \frac{1}{2}
  \left(m_u(2~\gev)+m_d(2~\gev) \right)\\
  &=3.5^{+0.7}_{-0.3}\ \mev
  \end{split}
\end{equation}
for nonstrange quarks and 
\begin{equation}
   m(2~\gev)=96^{+8}_{-4}\ \mev
\end{equation}
for strange quarks~\cite{PDG2017}.
In both~(\ref{running_coupling}) and~(\ref{running_light_quark}), we set $n_f=4$.

Renormalization-group arguments identify
our renormalization scale as $\mu = \tau^{1/4}$ \cite{Bertlmann1985,Orlandini2001}, putting a lower bound on our choice of $\tau$ restricted by the reliability of perturbation theory.  A related issue  associated with $\tau$ is reliability of the GSRs as quantified by the relative contributions of perturbative versus non-perturbative effects and the relative contributions of the resonance versus continuum. 
We therefore restrict our analysis to $\tau\geq M_{\tau}$, approximately equivalent to $\tau > 10\,\gev^4$ as discussed in Section~\ref{V}. 
We also work with an upper bound  $\tau\le 20\ \gev^4$ emerging from the H\"older inequality constraint \eqref{holder_coefficient}, as  presented in detail in Section~\ref{V}.

Turning to the condensates,
the value of the RG-invariant quantity $\langle m \overline{q}q\rangle$
is well-known from PCAC~\cite{GMOR1968}.  
Using the conventions of \cite{Narison2004}, we have
\begin{equation}\label{quarkcondensate}
  \langle m\overline{q}q\rangle =
  \begin{cases}
  -\frac{1}{2} f_{\pi}^2 m_{\pi}^2,\ \text{for nonstrange}\ q\\
  -\frac{1}{2} f_K^2 m_K^2,\ \text{for strange}\ q\ 
  \end{cases}
\end{equation}
where PDG values are used for the meson masses~\cite{PDG2017} 
and the decay constants are~\cite{Rosner2015} 
\begin{equation}
  f_{\pi} = (92.2\pm3.5)\ \mev\ ,\
  f_K = (110.0\pm4.2)\ \mev
  .
  \label{decay_couplings}
\end{equation}
We use the following value for the 4d gluon condensate~\cite{Narison2012}:
\begin{equation}
\gluonfourd = (0.075\pm0.020)\,\gev.
\label{gluoncondensate}
\end{equation}
The nonstrange- and strange-flavored 5d mixed condensates are estimated by \cite{Beneke1992,Belyaev1982} to be
\begin{equation}
   \frac{m\mixed}{\langle m\overline{q}q\rangle} 
   \equiv M^{2}_{0} 
   = (0.8 \pm 0.1)\,\gev^{2}.
 \label{mixedcondensate}
\end{equation}
Finally, we note that while the 6d quark and gluon condensates were included in the correlator calculation~(\ref{generalcorrelator}),
Table~\ref{A_coefficients} shows that neither contributes to the 
$k=0$ GSRs~(\ref{subtractedGSR}) or NGSRs~(\ref{gsrNormalized}).

\begin{figure*}[htbp]
  \centering
\includegraphics[width=\textwidth]{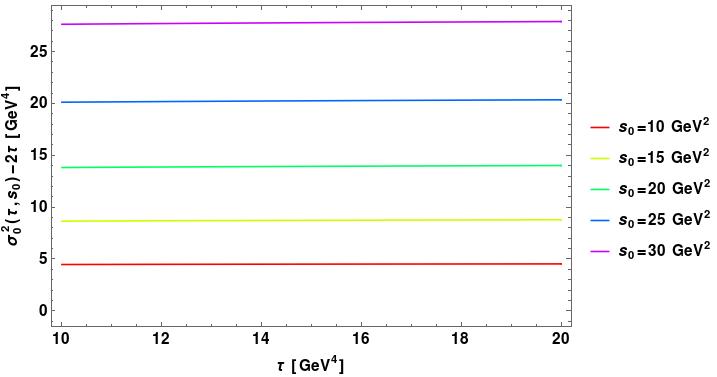}
\caption{\label{sigmaSquaredPlot} 
  The QCD prediction for the quantity $\sigma_0^2-2\tau$ (where $\sigma_0^2$ is defined in~(\ref{sigmaderek})) 
  for nonstrange quarks versus $\tau$ at several values of 
  the continuum threshold $s_0$.
  }
\end{figure*}

As noted in Section~\ref{III}, a SNR analysis of $0^{+-}$ light quarkonium hybrids 
fails within the LSRs methodology, and so we turn our attention to models with distributed resonance strength using GSRs. As confirmation of the consistency between the LSRs and GSRs methodology, we analysed the original stabilizing channels in the LSRs methodology $J^{PC}\in\{0^{\pm\pm},\,1^{\pm\pm}\}$ \cite{Govaerts1985} and found excellent agreement between the results for both mass predictions and continuum onsets.
To confirm the need for a distributed resonance model in the case of $J^{PC}=0^{+-}$, we consider the quantity~\cite{Harnett2001}
\begin{equation}\label{sigmaderek}
  \sigma^{2}_{0} (\tau,\,s_0) \equiv 
  \frac{M_{0,2}(\tau,\,s_0)}{M_{0,0}(\tau,\,s_0)} 
  - \left(\frac{M_{0,1}(\tau,\,s_0)}{M_{0,0}(\tau,\,s_0)}\right)^2
\end{equation}
where the QCD  moments, $M_{k,n}(\tau,\,s_0)$, were defined in~(\ref{gsrMoment}).
Combining~(\ref{masterFormulaGSR}) and~(\ref{sigmaderek}) gives
\begin{equation}\label{sigmaDuality}
\begin{split}
  \sigma_0^2 (\tau,\,s_0) = & \frac{\int_{t_0}^{\infty}(t^2+2\tau)\rho^{\text{had}}(t)\,\dt}%
                    {\int_{t_0}^{\infty}\rho^{\text{had}}(t)\,\dt}\\
    & - \left(\frac{\int_{t_0}^{\infty}t\rho^{\text{had}}(t)\,\dt}%
                 {\int_{t_0}^{\infty}\rho^{\text{had}}(t)\,\dt}\right)^2.
\end{split}
\end{equation}
For a SNR model, substituting~(\ref{singleNarrowResonance}) into~(\ref{sigmaDuality}) 
yields
\begin{equation}\label{sigmaSNRderek}
\sigma_{0}^{2}(\tau,\,s_0) = 2\tau.
\end{equation}
Hence, the quantity 
$\sigma_{0}^{2}(\tau,\,s_0) - 2\tau$ 
provides a QCD-calculated, model-independent way to 
assess the suitability of representing a particular hadronic spectral
function as a single narrow resonance.
If $\sigma_{0}^{2}(\tau,\,s_0) - 2\tau\approx0$, then
a single narrow resonance model is appropriate. 
On the other hand, if $\sigma_{0}^{2}(\tau,\,s_0) - 2\tau\not\approx 0$,
then the hadronic spectral function has distributed resonance strength.
And so, in Figure~\ref{sigmaSquaredPlot}, we plot the QCD prediction 
$\sigma_0^2(\tau,\,s_0)-2\tau$
versus $\tau$ for nonstrange quarks at several values of $s_0$ 
over the range $10\ \gev^2\leq s_0\leq 30\ \gev^2$.
Clearly, $\sigma_0^2(\tau,\,s_0)-2\tau\not\approx 0$, 
providing further motivation to consider models other than the SNR.
An analogous analysis for strange quarks leads to the same conclusion.

\begin{figure}[htbp]
  \centering
\includegraphics[width=\columnwidth]{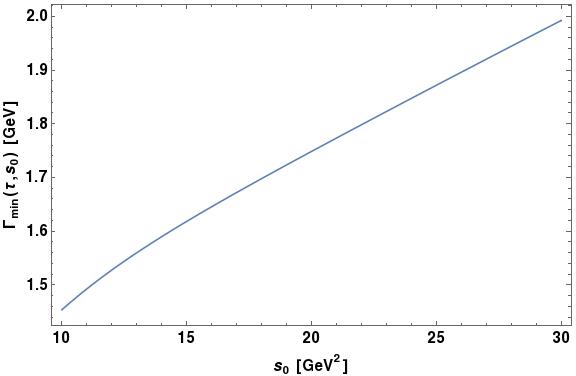}
\caption{\label{GammaMinPlot} Minimum rectangular pulse resonance width,
  $\Gamma_{\text{min}}$ from~(\ref{GammaMin}), at $\tau=10\ \gev^4$
  for nonstrange quarks versus the continuum threshold, $s_0$.}
\end{figure}

If the distributed resonance strength indicated by Figure~\ref{sigmaSquaredPlot}
is due to a single wide resonance (SWR), 
then we can determine a rough lower bound on the resonance's width  
using a rectangular pulse resonance model,
\begin{widetext}
\begin{equation}\label{pulsederek}
  \rho^\mathrm{had} \left(t\right)=\frac{\pi f}{2m_H\Gamma} \left[\theta\left( t-m_H^{2} 
  + m_H \Gamma \right) - \theta\left( t-m_H^{2} - m_H \Gamma \right)\right],
\end{equation}
\end{widetext}
where $f$ is the resonance's coupling, $\Gamma$ is its width, and $m_H$ is its mass. 
Substituting~(\ref{pulsederek}) into~(\ref{sigmaDuality}) gives 
\begin{gather}
  \sigma_0^2(\tau,\,s_0) = 2\tau + \frac{1}{3}m_H^2 \Gamma^2\label{sigmaRectangle}\\
  \implies \Gamma = \frac{1}{m_H}\sqrt{3\left(\sigma_0^2(\tau,\,s_0)-2\tau \right)}.
    \label{pulseGamma}
\end{gather}
From~(\ref{pulseGamma}), we see that $\Gamma$ decreases as $m_H$ increases.
However, to ensure that the resonance does not merge with the continuum, we require
\begin{equation}
  m_H^2 + m_H \Gamma < s_0
\end{equation}
which implies that the largest possible resonance mass for a particular $s_0$ is given by 
\begin{equation}
  m_{H,\,\text{max}}(\tau,\,s_0) = \sqrt{s_0-\sqrt{3(\sigma_0^2(\tau,\,s_0) - 2\tau)}}
\end{equation}
where we have used~(\ref{sigmaRectangle}).
By letting $m_H\rightarrow m_{H,\,\text{max}}$ in~(\ref{pulseGamma}), we find that
the smallest possible resonance width for a particular $s_0$ is given by
\begin{equation}\label{GammaMin}
  \Gamma_{\text{min}}(\tau,\,s_0) = \sqrt{\frac{3(\sigma_0^2(\tau,\,s_0)-2\tau)}%
  {s_0-\sqrt{3(\sigma_0^2(\tau,\,s_0)-2\tau)}}}.
\end{equation}
From Figure~\ref{sigmaSquaredPlot}, we see that $\sigma_0^2(\tau,\,s_0)-2\tau$ shows
almost no $\tau$-dependence; hence, the same can be said about $\Gamma_{\text{min}}(\tau,\,s_0)$.
In Figure~\ref{GammaMinPlot}, we plot $\Gamma_{\text{min}}(\tau,\,s_0)$ versus $s_0$ at $\tau = 10\ \gev^4$
for nonstrange quarks.  
An analogous plot for strange quarks looks nearly identical.
At $s_0=10\ \gev^2$, we find that $\Gamma_{\text{min}}\approx 1.46\ \gev$, far larger than a typical
hadron width.  As $s_0$ increases, so too does $\Gamma_{\text{min}}$.
For these reasons, we abandon SWR models in favour of a multi-resonance model.

\begin{figure}[htbp]
  \centering
\includegraphics[width=\columnwidth]{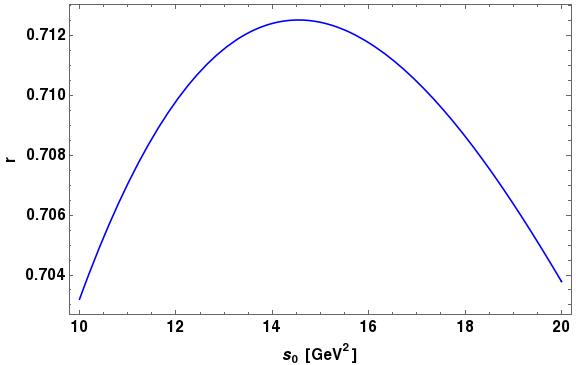}
\caption{\label{rPlot} Plot of the best fit $r$ (defined in~(\ref{rDefn}))
to~(\ref{gsrDNRDerek}) at $\tau=10\ \gev^4$ as a function of continuum threshold, $s_0$.}
\end{figure}

\begin{figure}[htbp]
  \centering
\includegraphics[width=\columnwidth]{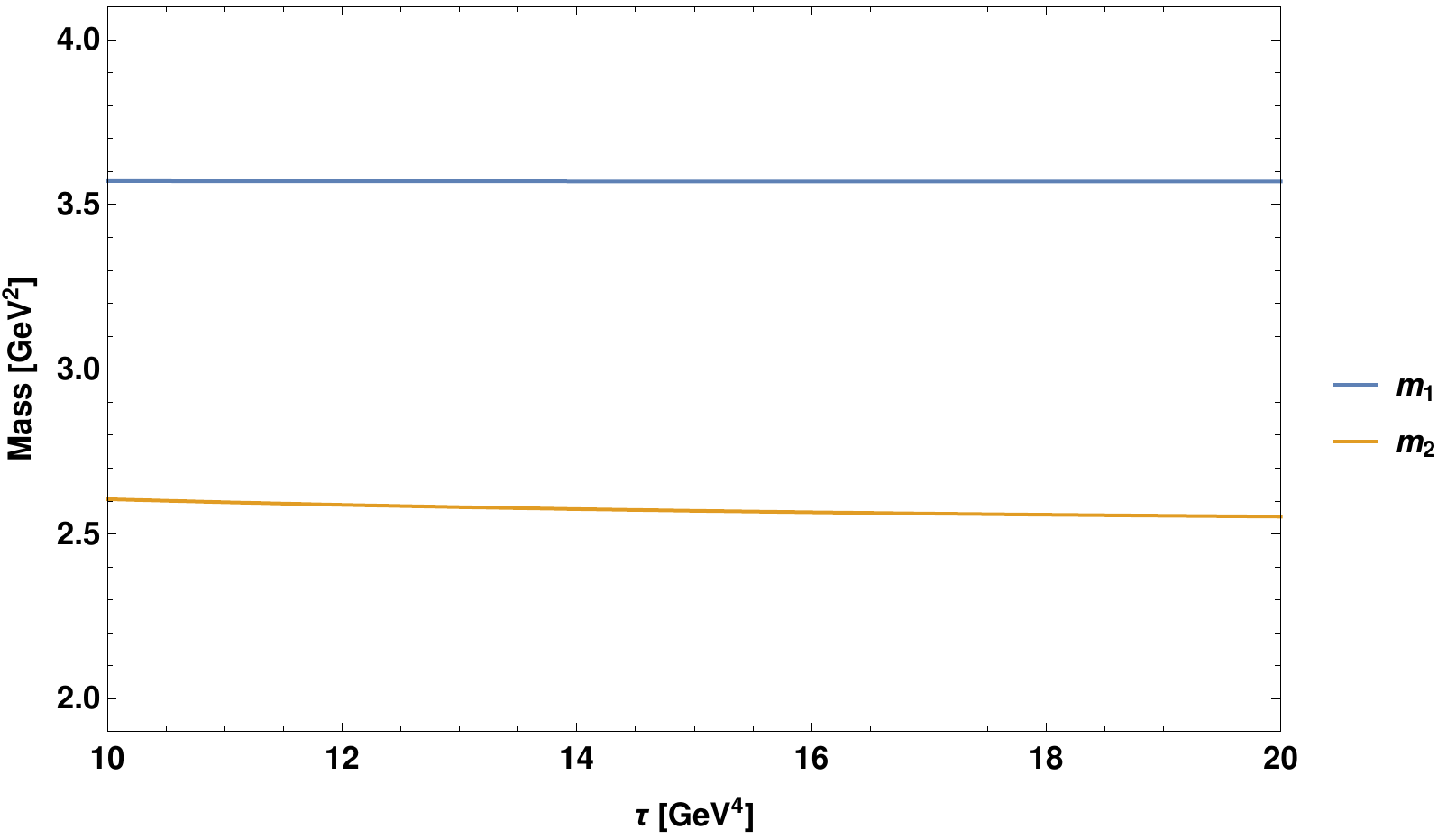}
\caption{Plot of $0^{+-}$ light quarkonium hybrid 
masses $m_{1}(\tau,\,s_0^\mathrm{opt})$ and $m_{2}(\tau,\,s_0^\mathrm{opt})$
of the DNR model~(\ref{doubleNarrowDerek})
at continuum threshold $s_0^\mathrm{opt} = 14.5\,\gev^{2}$.}
\label{fig.DNRmasses}
\end{figure}

\begin{figure*}[htbp]
  \centering
\includegraphics[width=\textwidth]{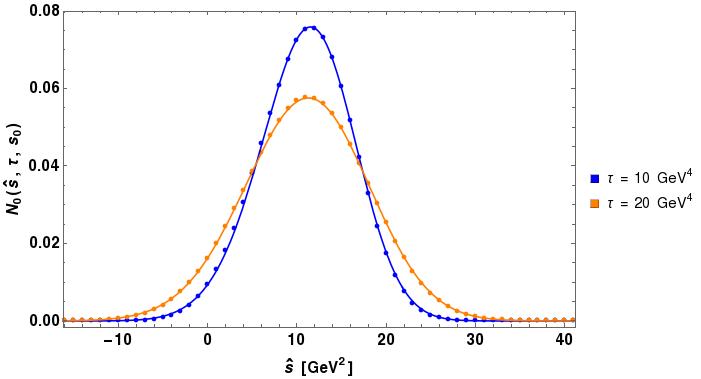}
\caption{Comparison of the two sides of~(\ref{gsrDNRDerek})
for nonstrange DNR parameters~(\ref{dnr.rDerek})--(\ref{dnr.m2Derek})
and for $\tau=10\ \gev^4$ and $\tau=20\ \gev^4$ 
at $s_{0}^\mathrm{opt} = 14.5\ \gev^2$.  Solid curves correspond to the left-hand side
of~(\ref{gsrDNRDerek}); dots correspond to the right-hand side at selected values of $\hat{s}$.}
\label{fig.modelcompare}
\end{figure*}

We consider a double narrow resonance (DNR) model
\begin{equation}\label{doubleNarrowDerek}
  \rho^\mathrm{had}(t) = \pi\Big(f_{1}^{2}\delta\left(t-m_{1}^2\right) 
    + f_{2}^{2}\delta\left(t-m_{2}^2\right)\Big),
\end{equation}
where $f_1,\,f_2$ and $m_1,\,m_2$ are the resonances' couplings and masses
respectively.  Substituting~(\ref{doubleNarrowDerek}) into~(\ref{masterFormulaNGSR}) gives
\begin{equation}\label{gsrDNRDerek}
  N_{0}\left( \hat{s},\,\tau,\,s_0\right)= \frac{\Bigg( r e^{-\frac{(\hat{s} - m_{1}^{2})^2}{4\tau}}
  +  (1-r) e^{-\frac{(\hat{s} - m_{2}^{2})^2}{4\tau}} \Bigg)}{\sqrt{4\pi \tau}}
\end{equation}
where
\begin{equation}\label{rDefn}
 r= \frac{f_{1}^{2}}{f_{1}^{2}+f_{2}^{2}}
 \Longleftrightarrow
 1-r = \frac{f_{2}^{2}}{f_{1}^{2}+f_{2}^{2}}.
\end{equation}
At fixed values of $\tau$ and $s_0$, we perform a fit of~(\ref{gsrDNRDerek}) over 
$\hat{s}$~\footnote{using the Mathematica v11 command NonlinearModelFit}
to find best fit parameters for $r$, $m_1$, and $m_2$.
In Figure~\ref{rPlot}, we plot the best fit $r$ versus $s_0$ at $\tau=10\ \gev^4$ 
for nonstrange quarks.
Again, an analogous plot for strange quarks looks nearly identical.
From the $s_0$-stability in $r$ versus $s_0$, 
we determine an optimized continuum onset for both the nonstrange- and strange-flavored cases as
\begin{equation}\label{dnr.s0}
s_0^{\mathrm{opt}} = (14.5 \pm 1.2)\ \gev^{2}
\end{equation}
where the uncertainties originate from the QCD input parameters; details of the uncertainty analysis are discussed below. 
Then, a fit to~(\ref{gsrDNRDerek}) at $s_0=14.5\ \gev^2$ and $\tau=10\ \gev^4$,
leads to DNR model parameters 
\begin{align}
\label{dnr.rDerek}
r & = 0.712\pm0.005\\
\label{dnr.m1Derek}
m_1 & = 3.57\pm0.15  \,\gev\\
\label{dnr.m2Derek}
m_2 & = 2.60\pm0.14  \,\gev
\end{align}
in the nonstrange-flavored case and
\begin{align}
\label{dnr.r}
r & = 0.711\pm0.005\\
\label{dnr.m1}
m_1 & = 3.57\pm0.13  \,\gev\\
\label{dnr.m2}
m_2 & = 2.60\pm0.14  \,\gev
\end{align}
in the strange-flavored case. Figure~\ref{fig.DNRmasses} shows negligible $\tau$ dependence in the mass predictions.
Figure~\ref{fig.modelcompare} shows comparisons between the
the NGSRs and the DNR model 
(respectively the left- and right-hand sides of~(\ref{gsrDNRDerek}))
for parameters~(\ref{dnr.rDerek})-(\ref{dnr.m2Derek}) 
at $\tau = 10\ \gev^{4}$ and $\tau = 20\ \gev^{4}$.
We note that the strange and nonstrange $0^{+-}$ hybrid mass predictions are degenerate within the uncertainties of our analysis; we find this to be consistent with other recent SR analyses \cite{Ho2017,Chen2017}. We note that the correlator terms that contain the strange quark mass and condensates are numerically  small in our calculation, and do not significantly impact the resulting mass prediction.
The relatively small numerical difference between strange and non-strange $0^{+-}$ hybrids could suggest a dominance of constituent gluonic effects in these systems.


\begin{figure}[htbp]
  \centering
\includegraphics[width=\columnwidth]{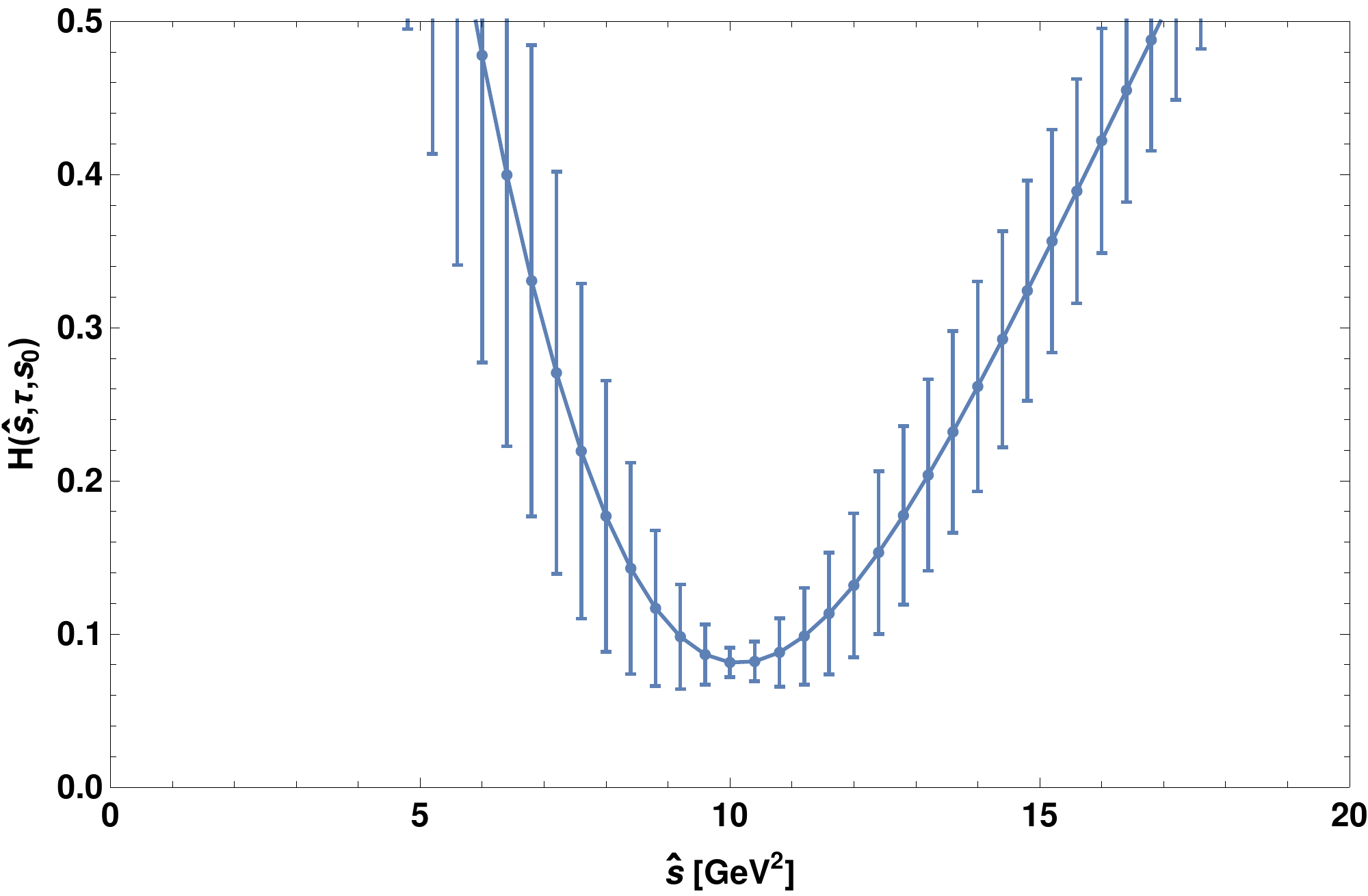}
\caption{Plot of inequality test \eqref{holder_coefficient} for $\tau = 10\ \gev^{4}$ with error bars displayed. Errors are due to variations in the condensate parameters, variations in $\alpha_s$, and uncertainties in $s_0^\mathrm{opt}$ \eqref{dnr.s0}.
\label{fig.holdererror}}
\end{figure}

Utilizing the H\"older Inequality test~\eqref{holder_coefficient}, 
we can perform a consistency check on our analysis. To determine whether~\eqref{holder_coefficient} 
is satisfied within the expected uncertainties of the GSRs, 
we examine the inequality at $s_0^\mathrm{opt} = 14.5\ \gev^{2}$ for various values of $\tau$.  
Because our QCD calculations of Wilson coefficients are truncated perturbative series in $\alpha_s$, 
in addition to the QCD parameter uncertainties, we use the $1^{-+}$ channel~\cite{Jin2001}
to provide an estimated next-order perturbative correction characteristic of hybrid correlators.
We find that the H\"older inequality constraint~\eqref{holder_coefficient} 
is violated for $\tau\gtrsim20\ \gev^{4}$, and the inequality test for the minimum value $\tau=10\,\gev^4$ is shown in Figure~\ref{fig.holdererror}.   
Thus, the $\tau$ range used in our analysis, $10\ \gev^4<\tau<20\ \gev^4$,  
is consistent with the H\"older inequality.

To explore the lower bound on $\tau$ in more detail, we consider OPE convergence and  resonance dominance in the GSR.  As in LSRs, a reliable GSR analysis requires that perturbation theory dominates power-law corrections and that the resonance  contributions dominate the continuum. The average relative contribution of the non-perturbative terms is calculated over the region $10\,\rm{GeV^2}- \sqrt{2\tau} <\hat s< 10\,\rm{GeV^2}+ \sqrt{2\tau}$ to encompass the peak in Figure \ref{fig.modelcompare}.  For $\tau=10\,{\rm GeV^4}$, the $\hat s$-averaged non-perturbative contributions are less than 20\% of the total and are thus safely controlled.  As $\tau$ decreases, the relative non-perturbative contribution increases  (e.g., to 25\% at  $\tau=5\,{\rm GeV^4}$).  The relative contribution of the resonance versus continuum contribution is much more sensitive to $\tau$. For  $\tau=10\,{\rm GeV^4}$ the $\hat s$-averaged ratio of resonance to continuum effects is 50\%  but for   $\tau=5\,{\rm GeV^4}$  the ratio decreases to 30\%.  We thus conclude that the criteria of OPE convergence and  resonance dominance  requires $\tau>10\,{\rm GeV^4}$ for a reliable GSR analysis. The combination of the H\"older inequality, OPE convergence, and resonance dominance constrains our GSR window of analysis to $10\ \gev^4<\tau<20\ \gev^4$.

We verify the $s_0$ optimization~(\ref{dnr.s0}) obtained from Figure \ref{rPlot} by looking at an independent analysis developed in~\cite{Harnett2001} 
based on the properties of the $\hat s$ peak position (maximum) of the NGSRs. For the SNR 
model~\eqref{singleNarrowResonance} the $\hat s$-peak occurs at $\hat s=m^2$, independent of $\tau$.  
Thus, an alternative $s_0$-optimization criterion for the SNR is minimizing the $\tau$-dependence of the peak position $\hat s_{\mathrm{peak}}\left(\tau,\, s_0\right)$ defined implicitly from 
\begin{equation}\label{peakdrift}
\biggr.\frac{\partial}{\partial\hat{s}}N^{\mathrm{QCD}}_{0}\left( \hat s,\,\tau,\,s_0\right)\biggr|_{\hat s=\hat{s}_{\mathrm{peak}}\left(\tau,\,s_0\right)} = 0.
\end{equation}
For the DNR model~\eqref{doubleNarrowDerek}, the peak position acquires $\tau$-dependence modeled by
\begin{equation}
\hat s_{\mathrm{peak}}\left(\tau,s_0\right)=A + \frac{B}{\tau}+\frac{C}{\tau^2}+\frac{D}{\tau^3}
\end{equation}
where the unknown parameters $\{A,B,C,D\}$ are constrained by minimizing the $\chi^2$
\begin{widetext}
\begin{equation}\label{chisquared.peakdriftDerek}
\chi^{2}\left(A,B,C,D, s_0\right) = \sum_{\tau=10\,\gev^4}^{20\,\gev^4}\left( \frac{A + \frac{B}{\tau}+\frac{C}{\tau^2}+\frac{D}{\tau^3}}{\hat{s}_{\mathrm{peak}}\left(\tau,s_0\right)}-1 \right)^{2}.
\end{equation}
\end{widetext}
By minimizing (\ref{chisquared.peakdriftDerek}) with respect to $A$, $B$, $C$, $D$, and $s_0$, we find an optimum continuum threshold $s_{0}^\mathrm{opt} = 14.0 \,\gev^{2}$ 
in excellent agreement with the value obtained in~(\ref{dnr.s0}).

To obtain errors in $s_0^\mathrm{opt}$, $r$, $m_1$, and $m_2$, we examine how the errors in the QCD parameters impact the values of these optimized parameters by varying each independently and examining the impact on the model parameters. Additionally, there exists a methodological error in determining $s_0^\mathrm{opt}$ as the variance in the QCD parameters will affect the stability point of $r$. Contributions to the error in $s_0^\mathrm{opt}$ are summarized in Table \ref{table.error.s0} and contributions to the error in the DNR model parameters are summarized in Tables \ref{table.error.r}-\ref{table.error.m2}. The dominant error in $s_0^\mathrm{opt}$ comes from the variation in $\gluonfourd$; in determining errors in the DNR parameters, the error in $r$ is driven by the variation in $\gluonfourd$ while the dominant errors in the masses $m_1$ and $m_2$ arise from variations in $s_0^\mathrm{opt}$, followed by $\gluonfourd$. Errors in $\quarkthreed$ and $\mixed$ contribute negligibly in the error of all DNR parameters. Adding the values summarized in Tables \ref{table.error.s0}-\ref{table.error.m2} in quadrature gives us a conservative error estimate summarized in Table \ref{table.error.summary}; as the driving errors in each parameter are approximately equivalent for the upper and lower bounds of the corresponding QCD parameters, we express our DNR parameters (\ref{dnr.s0})-(\ref{dnr.m2}) with symmetric error, taking the most conservative bound.
\begin{table*}[h]
\centering
\caption{Contributions to $s_0^\mathrm{opt}$ error at $\tau = 10\,\gev^{4}$ due to variations in QCD parameter error. Columns $\pm\delta$ indicate variations in DNR parameters at the upper ($+\delta$) and lower ($-\delta$) bounds of the corresponding QCD parameters. }
\label{table.error.s0}
\begin{tabular}{|c|c|c|c|c|}
\hline
\multirow{ 2}{*}{Error Source} &  \multicolumn{2}{|c|}{Nonstrange} &  \multicolumn{2}{|c|}{Strange} \\ \cline{2-5} 
 & $+\delta$ & $-\delta$ & $+\delta$ & $-\delta$   \\ \hline
$\mquarkthreed$ & $5.62\times 10^{-3}$ & $-6.74 \times 10^{-4}$ & $9.38 \times 10^{-3}$& $3.09 \times 10^{-3}$ \\ \hline
$\gluonfourd$ & $-9.20\times 10^{-1}$ & $1.18\times 10^{0}$ & $-9.16 \times 10^{-1}$ & $1.18 \times 10^{0}$ \\ \hline
$\mixed$ & $1.11\times 10^{-4}$ & $1.00\times 10^{-3}$ & $3.66 \times 10^{-3}$ & $4.77 \times 10^{-3}$ \\ \hline
$\alpha_s$ & $ 1.70 \times 10^{-1}$ & $ -1.86 \times 10^{-1}$ & $1.74 \times 10^{-1}$ & $ -1.82 \times 10^{-1} $ \\ \hline
\end{tabular}
\end{table*}

\begin{table*}[h]
\centering
\caption{Contributions to $r$ error at $\tau = 10\,\gev^{4}$ due to variations in QCD parameter error. Columns $\pm\delta$ indicate variations in DNR parameters at the upper ($+\delta$) and lower ($-\delta$) bounds of the corresponding QCD parameters. }
\label{table.error.r}
\begin{tabular}{|c|c|c|c|c|}
\hline
\multirow{ 2}{*}{Error Source} &  \multicolumn{2}{|c|}{Nonstrange} &  \multicolumn{2}{|c|}{Strange} \\ \cline{2-5} 
 & $+\delta$ & $-\delta$ & $+\delta$ & $-\delta$   \\ \hline
$\mquarkthreed$ & $-2.86\times 10^{-6}$ & $2.76 \times 10^{-6}$ & $-2.58 \times 10^{-6}$ & $3.04 \times 10^{-6}$ \\ \hline
$\gluonfourd$ & $4.79\times 10^{-3}$ & $-4.66\times 10^{-3}$ & $4.79 \time 10^{-3}$ & $-4.66 \times 10^{-3}$ \\ \hline
$\mixed$ & $1.43\times 10^{-6}$ & $-1.34\times 10^{-6}$  & $1.70 \times 10^{-6}$ & $-1.06 \times 10^{-6}$ \\ \hline
$\alpha_s$ & $ -7.93 \times 10^{-4}$ & $ 8.73 \times 10^{-4}$  & $-7.93 \times 10^{-4}$ & $8.73\times 10^{-4}$ \\ \hline
$s_0^\mathrm{opt} $ & $ 5.09 \times 10^{-4} $  & $ 3.53\times 10^{-4} $& $3.57 \times 10^{-4}$ & $4.81 \times 10^{-4}$ \\ \hline
\end{tabular}
\end{table*}

\begin{table*}[h]
\centering
\caption{Contributions to $m_1$ error at $\tau = 10\,\gev^{4}$ due to variations in QCD parameter error. Columns $\pm\delta$ indicate variations in DNR parameters at the upper ($+\delta$) and lower ($-\delta$) bounds of the corresponding QCD parameters. }
\label{table.error.m1}
\begin{tabular}{|c|c|c|c|c|}
\hline
\multirow{ 2}{*}{Error Source} &  \multicolumn{2}{|c|}{Nonstrange} &  \multicolumn{2}{|c|}{Strange} \\ \cline{2-5} 
 & $+\delta$ & $-\delta$ & $+\delta$ & $-\delta$   \\ \hline
$\mquarkthreed$ & $-6.84\times 10^{-6}$ & $6.58 \times 10^{-6}$   & $-7.06\times 10^{-6}$ & $6.35\times 10^{-6}$ \\ \hline
$\gluonfourd$ & $5.79\times 10^{-3}$ & $-7.02\times 10^{-3}$   & $5.79\times 10^{-3}$ & $-7.02\times 10^{-3}$ \\ \hline
$\mixed$ & $-3.07\times 10^{-6}$ & $3.01\times 10^{-6}$   & $-3.29\times 10^{-6}$ & $2.79\times 10^{-6}$ \\ \hline
$\alpha_s$ & $ -1.06 \times 10^{-3}$ & $ 1.13 \times 10^{-3}$   & $-1.06\times 10^{-3}$ & $1.13\times 10^{-3}$  \\ \hline
$s_0^\mathrm{opt} $ & $ 1.48 \times 10^{-1} $  & $ -1.21\times 10^{-1} $   & $1.49\times 10^{-1}$ & $-1.20\times 10^{-1}$ \\ \hline
\end{tabular}
\end{table*}

\begin{table*}[h]
\centering
\caption{Contributions to $m_2$ error at $\tau = 10\,\gev^{4}$ due to variations in QCD parameter error. Columns $\pm\delta$ indicate variations in DNR parameters at the upper ($+\delta$) and lower ($-\delta$) bounds of the corresponding QCD parameters. }
\label{table.error.m2}
\begin{tabular}{|c|c|c|c|c|}
\hline
\multirow{ 2}{*}{Error Source} &  \multicolumn{2}{|c|}{Nonstrange} &  \multicolumn{2}{|c|}{Strange} \\ \cline{2-5} 
 & $+\delta$ & $-\delta$ & $+\delta$ & $-\delta$   \\ \hline
$\mquarkthreed$ & $-3.72\times 10^{-5}$ & $3.58 \times 10^{-5}$   & $-3.79\times 10^{-5}$ & $3.51\times 10^{-5}$  \\ \hline
$\gluonfourd$ & $2.81\times 10^{-2}$ & $-3.64\times 10^{-2}$   & $2.81\times 10^{-2}$ & $-3.64\times 10^{-2}$  \\ \hline
$\mixed$ & $-2.01\times 10^{-5}$ & $1.99\times 10^{-5}$   & $-2.08\times 10^{-5}$ &$1.93\times 10^{-5}$   \\ \hline
$\alpha_s$ & $ -5.36 \times 10^{-3}$ & $ 5.63 \times 10^{-3}$   & $-5.36\times 10^{-3}$ & $5.63\times 10^{-3}$   \\ \hline
$s_0^\mathrm{opt} $ & $ 1.40 \times 10^{-1} $  & $ -1.12\times 10^{-1} $   & $1.40\times 10^{-1}$ & $-1.11\times 10^{-1}$  \\ \hline
\end{tabular}
\end{table*}

\begin{table*}[h]
\centering
\caption{Calculated total errors in $s_0^\mathrm{opt}$, $r$, $m_1$, $m_2$ from contributions in Tables \ref{table.error.s0} - \ref{table.error.m2}.}
\label{table.error.summary}
\begin{tabular}{|c|c|c|}
\hline
\multirow{2}{*}{Parameter} & \multicolumn{2}{|c|}{Value} \\ \cline{2-3}
& Nonstrange & Strange  \\ \hline
$s_0^\mathrm{opt}$ & $14.5^{+1.2}_{-0.9}$ &$14.5^{+1.2}_{-0.9}$\\ \hline
$r$ & $ 0.712\pm0.005$& $0.711\pm0.005$\\ \hline
$m_1$ & $3.57^{+0.15}_{-0.12}$& $3.57^{+0.13}_{-0.12}$ \\ \hline
$m_2$ & $2.60^{+0.14}_{-0.12}$& $2.60^{+0.14}_{-0.12}$\\ \hline
\end{tabular}
\end{table*}

\section{Discussion}\label{VI}
We have calculated 5d and 6d
QCD condensate contributions to all spin-0 and spin-1 light quarkonium
hybrid correlators with the goal of obtaining QCD LSRs 
mass predictions in the previously-unstable channels of~\cite{Govaerts1985}.  
However, 
the 6d gluon and quark condensate contributions do not have an imaginary part 
and hence do not contribute to the LSRs. 
Also, the 5d mixed condensate contributions turn out to be small.
We therefore focused on the suggestion of References~\cite{Govaerts1985,Braun1986} 
 that a distribution of resonance strength could be the source of instability, 
 a scenario ideally suited to GSRs
methods~\cite{Bertlmann1985,Orlandini2001,Harnett2001}. 
The $0^{+-}$ channel was chosen for detailed investigation because 
of its phenomenological significance in light of the GlueX experiment.
Furthermore, a model-independent analysis of the $0^{+-}$ hadronic spectral function implies that there 
is a distribution of resonance strength in this channel.

In examining the  SWR~(\ref{pulsederek}) and DNR~(\ref{doubleNarrowDerek}) models, 
we found that the DNR model provided excellent agreement between QCD and phenomenology.
(See Figure~\ref{fig.modelcompare}.)  The SWR model was
rejected on the basis of an atypically large resonance width. 
In the DNR model, we find degenerate predictions in the case of both nonstrange and strange quark states from the $0^{+-}$ current: a $2.60\pm0.14\,\gev$ state ($2.60\pm0.14\,\gev$ in the strange case) with $29\%$ relative coupling, and a state at $3.57\pm0.15\,\gev$ ($3.57\pm0.13\,\gev$) with $71\%$ relative coupling. The smaller coupling of the light state suggests the possibility of mixing with a tetraquark because the expected tetraquark mass range is above $2\,\gev$~\cite{Du2013}.

The lighter state is consistent with recent lattice results that find a predominantly
nonstrange state around $2.4\ \gev$ and a predominantly strange state around $2.5\ \gev$ in the $0^{+-}$ channel with $m_\pi =  391\ \mev$~\cite{Dudek2013}. 
Our lighter-state mass determination is somewhat larger than the 2.1--2.5~$\gev$ range of central values in~\cite{Braun1986}.
The literature does not provide a clear interpretation of the heavier $0^{+-}$ state; 
however lattice studies~\cite{Dudek2013}
of the lightest hybrid meson supermultiplet suggest that the $0^{+-}$ state exists as part of an excited hybrid supermultiplet with radially-excited $q\overline{q}$ pair 
(i.e., quark total angular momentum $L_{q\overline{q}} = 1$). 
We suggest that this heavier second state arising in the GSRs is a manifestation of an excited hybrid state.

In conclusion, we investigated light quarkonium,
exotic ($J^{PC} = 0^{+-}$) hybrid mesons with SWR and DNR models using a GSRs analysis. 
We disfavoured the SWR model as the predicted resonance width was too large. 
The double-narrow resonance model yielded two $0^{+-}$ hybrid states:
$(2.60\pm0.14)\ \gev$ and $(3.57\pm0.15)\ \gev$ 
($(2.60\pm0.14)\ \gev$ and $(3.57\pm0.13)\ \gev$ in the strange case). 
Additionally, we explored using the H\"older inequality derived for the GSRs as a consistency check on our analysis. 
\section*{Acknowledgments}
We are grateful for financial support from the Natural Sciences and 
Engineering Research Council of Canada (NSERC), and the Chinese National Youth Thousand Talents Program.
\clearpage
\bibliographystyle{h-physrev}
\bibliography{lighthybrids}
\end{document}